\documentclass[letterpaper,twocolumn,english,aps,jcp,superscriptaddress,floatfix]{revtex4}
\usepackage[LGR,LGR,LGR,LGR,T1]{fontenc}
\usepackage[latin9]{inputenc}
\usepackage{color}
\usepackage{units}
\usepackage{amsmath}
\usepackage{amssymb}
\usepackage{graphicx}
\usepackage{esint}

\makeatletter

\@ifundefined{textcolor}{}
{%
 \definecolor{BLACK}{gray}{0}
 \definecolor{WHITE}{gray}{1}
 \definecolor{RED}{rgb}{1,0,0}
 \definecolor{GREEN}{rgb}{0,1,0}
 \definecolor{BLUE}{rgb}{0,0,1}
 \definecolor{CYAN}{cmyk}{1,0,0,0}
 \definecolor{MAGENTA}{cmyk}{0,1,0,0}
 \definecolor{YELLOW}{cmyk}{0,0,1,0}
}

\@ifundefined{definecolor}{\@ifundefined{definecolor}{\@ifundefined{definecolor}{\@ifundefined{definecolor}
 {\usepackage{color}}{}
}{}}{}}{}\renewcommand{\[}{\begin{equation}}

\renewcommand{\]}{\end{equation}}

\usepackage{babel}

\makeatother

\usepackage{babel}
\begin{document}
\global\long\def\avg#1{\langle#1\rangle}%
\global\long\def\p{\prime}%
\global\long\def\dg{\dagger}%
\global\long\def\ket#1{|#1\rangle}%
\global\long\def\bra#1{\langle#1|}%
\global\long\def\proj#1#2{|#1\rangle\langle#2|}%
\global\long\def\inner#1#2{\langle#1|#2\rangle}%
\global\long\def\tr{\mathrm{tr}}%
\global\long\def\pd#1#2{\frac{\partial#1}{\partial#2}}%
\global\long\def\spd#1#2{\frac{\partial^{2}#1}{\partial#2^{2}}}%
\global\long\def\der#1#2{\frac{d#1}{d#2}}%
\global\long\def\im{\imath}%
\global\long\def\onlinecite#1{\cite{#1}}%

\title{Metal adsorbate interactions and the convergence of density functional
calculations}
\author{Christoph Rohmann}
\address{Biophysics Group, Microsystems and Nanotechnology Division, Physical
Measurement Laboratory, National Institute of Standards and Technology,
Gaithersburg, MD 20899, USA}
\address{Maryland NanoCenter, University of Maryland, College Park, Maryland
20783, USA}
\author{Maicol A. Ochoa}
\address{Biophysics Group, Microsystems and Nanotechnology Division, Physical
Measurement Laboratory, National Institute of Standards and Technology,
Gaithersburg, MD 20899, USA}
\address{Maryland NanoCenter, University of Maryland, College Park, Maryland
20783, USA}
\author{Michael Zwolak}
\email{mpz@nist.gov}

\address{Biophysics Group, Microsystems and Nanotechnology Division, Physical
Measurement Laboratory, National Institute of Standards and Technology,
Gaithersburg, MD 20899, USA}
\date{\today}
\begin{abstract}
The adsorption of metal atoms on nanostructures, such as graphene
and nanotubes, plays an important role in catalysis, electronic doping,
and tuning material properties. Quantum chemical calculations permit
the investigation of this process to discover desirable interactions
and obtain mechanistic insights into adsorbate behavior, of which
the binding strength is a central quantity. Binding strengths, however,
vary widely in the literature, even when using almost identical computational
methods. To address this issue, we investigate the adsorption of a
variety of metals onto graphene, carbon nanotubes, and boron nitride
nanotubes. As is well-known, calculations on periodic structures require
a sufficiently large system size to remove interactions between periodic
images. Our results indicate that there are both direct and indirect
mechanisms for this interaction, where the latter can require even
larger system sizes than typically employed. The magnitude and distance
of the effect depends on the electronic state of the substrate and
the open- or closed-shell nature of the adsorbate. For instance, insulating
substrates (e.g., boron nitride nanotubes) show essentially no dependence
on system size, whereas metallic or semi-metallic systems can have
a substantial effect due to the delocalized nature of the electronic
states interacting with the adsorbate. We derive a scaling relation
for the length dependence with a representative tight-binding model.
These results demonstrate how to extrapolate the binding energies
to the isolated-impurity limit. 
\end{abstract}
\maketitle
Graphene, carbon nanotubes (CNTs), and boron nitride nanotubes (BNNTs)
have exceptional mechanical, thermal, and electronic properties. These
materials are thus the subject of intense research. Adsorption studies
range from hydrogen and fluorine to metals of the 3d, 4d, and 5d series
\citep{giesbers_interface-induced_2013,hong_evidence_2012,pasti_atomic_2018,dimakis_density_2017,manade_transition_2015,habenicht_adsorption_2014,liu_metals_2012,zhang_electrically_2012,nakada_migration_2011,valencia_trends_2010,sargolzaei_magnetic_2011,ding_engineering_2011,santos_magnetism_2010,zolyomi_first_2010,hu_density_2010,rohmann_interaction_2016,zhang_structural_2010,rohmann_interaction_2018}.
Among the latter are many density functional theory (DFT) studies
examining the behavior of single metal atoms. These, however, show
variation in the binding strengths up to several electron volts \citep{manade_transition_2015}.
In part, this is due to differences in the methods employed, such
as spin-polarized versus non-spin-polarized calculations, ultrasoft
pseudo-potentials versus projector augmented wa\textcolor{black}{ve
methods (PAWs), or the use of LDA versus GGA exchange functionals.
However, even studies employing almost identical techniques yield
different results. For example, comparing the investigations of Manadé
et al. \citep{manade_transition_2015}, Pašti et al. \citep{pasti_atomic_2018}
and Liu et al. \citep{liu_metals_2012} with respect to 3d metal adsorption,
which all employed the same computational package, the Perdew--Burke--Ernzerhof
Generalized-Gradient Approximation (PBE-GGA) exchange correlation
functional, and PAWs, as well as a $4\times4$ graphene super cell,
there are differences of up to 0.53 eV. From the information available,
these studies differ in energy cut offs (450 eV to 600 eV), $k$-point
meshes ($6\times6\times1$ to $10\times10\times1)$, and the graphene
layer spacings (1.5 nm to 2.0 nm). Although there are some differences
in the approaches, the differences in binding energy are unexpectedly
high given that DFT is regarded as well suited to study the effect
of single atom adsorption.}

\begin{figure}
\centering{}\includegraphics[width=0.99\columnwidth]{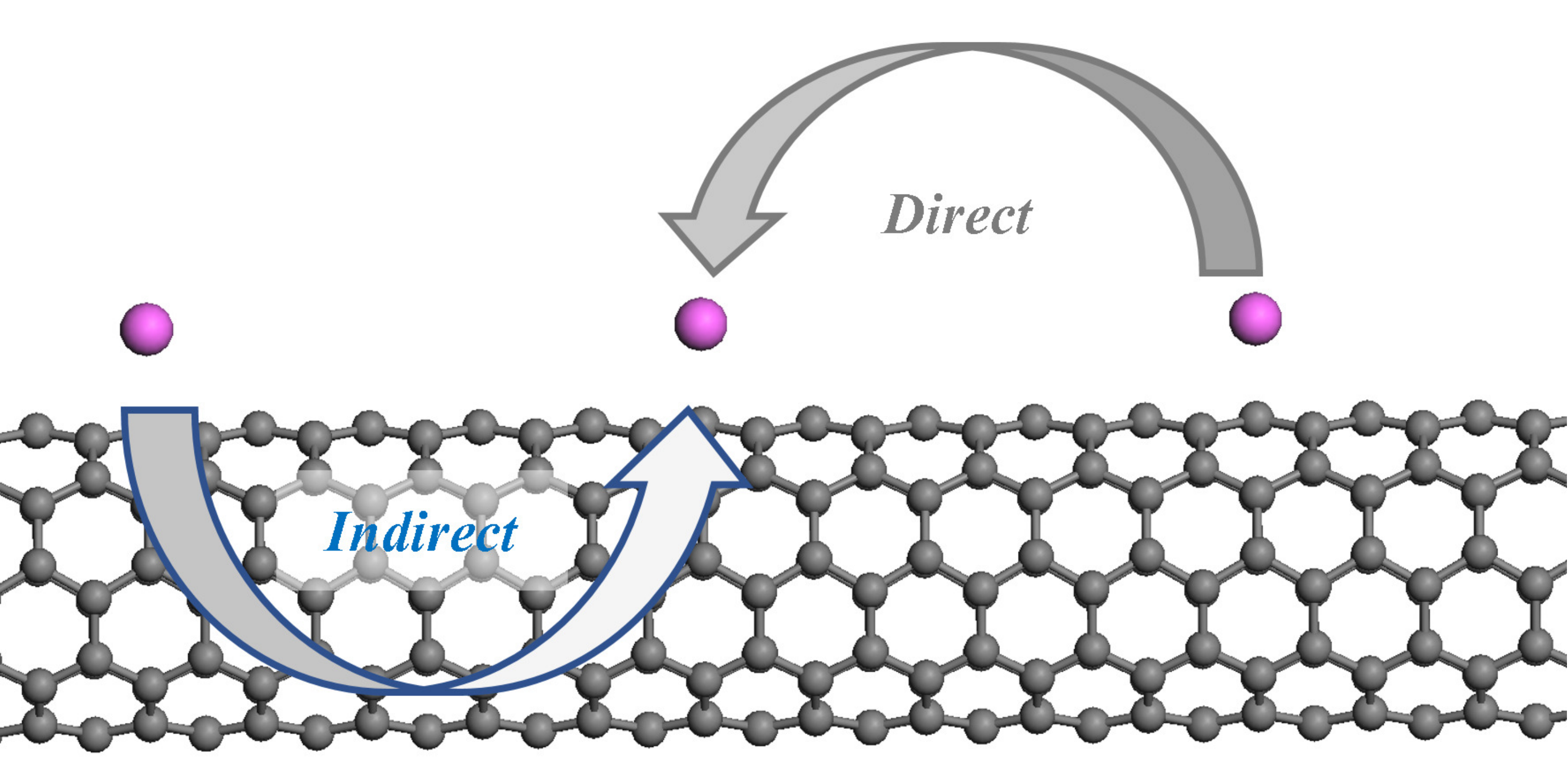}
\caption{\textbf{Metal adsorption on a CNT.} Within DFT calculations, adsorbates
``feel'' their periodic images through both direct (via free space)
and indirect (via the substrate) interactions. The latter can be substantial
and have a long range.\label{fig:Schematic}}
\end{figure}

Here, we investigate the binding energy dependence on the system size
for various 3d metals on graphene, CNTs, and BNNTs by means of DFT.
We aim to identify the cell size required to obtain, or extrapolate
to, the isolated impurity limit, as well as understand related sources
of error. Upon examining several metals on insulating, semiconducting,
semi-metallic, and conducting substrates, we observe a few classes
of behavior. Some combinations of metals and substrates yield a slow
decay or oscillatory behavior out to large system sizes. Other combinations
show little-to-no cell-size dependence, as is consistently seen with
metals on BNNTs. The use of different isolated atom calculations can
also influence results, as can the size of the vacuum gap. By examining
a one-dimensional tight-binding model, we demonstrate that the decay
in binding strength, including decay of oscillations, behaves as the
square of the inverse system size. The algebraic decay behaves as
the square of the inverse rather than just the inverse due to periodicity.
This scaling can be employed as a fitting form for more complex DFT
calculations. These results provide insight into the adsorption behavior
of metals on various substrates \ref{fig:Schematic}, as well as the
errors incurred in common computational methods.

\noindent \textbf{\textsc{Methodology}}

We employ spin-polarized DFT calculations using the Vienna Ab initio
Simulation Package (VASP) \citep{kresse_efficient_1996} and the PAW
method \citep{kresse_ultrasoft_1999} with the PBE-GGA functional.
\citep{perdew_generalized_1996}. All structures are relaxed until
the total energy converges to within $10^{-4}$ eV during the self-consistent
loop, with forces conver\textcolor{black}{ged to 0.1 eV/nm.{} An
energy cut-off of 450 eV is used for all calculations. Van der Waals
interactions are accounted for by the Grimme (D2) scheme \citep{grimme_semiempirical_2006}.
The valence electron configuration for each metal considered are Al:2s}\textsuperscript{\textcolor{black}{2}}\textcolor{black}{2p}\textsuperscript{\textcolor{black}{1}}\textcolor{black}{,
Ti:3d}\textsuperscript{\textcolor{black}{3}}\textcolor{black}{4s}\textsuperscript{\textcolor{black}{1}}\textcolor{black}{,
Fe:3d}\textsuperscript{\textcolor{black}{7}}\textcolor{black}{4s}\textsuperscript{\textcolor{black}{1}}\textcolor{black}{,
V:3p}\textsuperscript{\textcolor{black}{6}}\textcolor{black}{3d}\textsuperscript{\textcolor{black}{4}}\textcolor{black}{4s}\textsuperscript{\textcolor{black}{1}}\textcolor{black}{,
Ni:3d}\textsuperscript{\textcolor{black}{9}}\textcolor{black}{4s}\textsuperscript{\textcolor{black}{1}}\textcolor{black}{,
and Cu:3d}\textsuperscript{\textcolor{black}{10}}\textcolor{black}{4s.
The Methfessel-Paxton method with a smearing of 0.2 eV is taken for
metal adsorption to metallic CNTs and graphene, whereas a Gaussian
smearing of 0.02 eV is taken for semiconducting CNTs and insulating
BNNTs.}

\textcolor{black}{Prior to creating the supercells for the adsorption
studies, the unit cell of each nanotube (NT) and the graphene sheet
are optimized. The calculated parameters for the growth direction
(}\textit{\textcolor{black}{c}}\textcolor{black}{) of the nanotubes
are 0.252 nm for the (5,5) BNNT, 0.247 nm for the (5,5) CNT, and 0.428
nm for the (8,0) CNT, respectively. In the case of graphene, the }\textit{\textcolor{black}{a}}\textcolor{black}{{}
and}\textit{\textcolor{black}{{} b}}\textcolor{black}{{} lattice
parameters are found to be 0.247 nm. In all models, a vacuum layer
of at least 1.5 nm is added in }\textit{\textcolor{black}{a}}\textcolor{black}{{}
and }\textit{\textcolor{black}{b}}\textcolor{black}{{} for the NTs
and in }\textit{\textcolor{black}{c}}\textcolor{black}{{} for graphene.
To model the clean and adsorbate-covered structures, we examine nanotube
system sizes up to 2.56 nm (for the (8,0) CNT adsorbing Al, system
sizes go up to 3.84 nm while they are 3.42 nm in case of Ti and V),
while the dimensions of the graphene system go up to $2.47$ nm $\times$
$2.47$ nm. Table S1 (see supplementary material) lists the dimensions
of nanotubes and graphene sheets in this work.}

\textcolor{black}{We perform extensive testing to obtain the required
Monkhorst-Pack mesh \citep{monkhorst_special_1976}{} with the smallest
cell for each system. In case of the (5,5) BNNT, this represents the
$1\times1\times3$ supercell (0.741 nm) where a mesh of $1\times1\times2$
is found to be sufficient, while the $1\times1\times3$ supercell
}\linebreak{}
 \textcolor{black}{{} (0.755 nm) of the (5,5) CNT requires a $1\times1\times10$
mesh. In case of the (8,0) CNT, we employ a $1\times1\times2$ supercell
(0.854 nm) where a $k$-point mesh of $1\times1\times5$ is required,
with the exception of Al and V, which require a $1\times1\times7$
and $1\times1\times8$ mesh for convergence, respectively. Graphene
in a $3\times3$ supercell (0.740 nm $\times$ 0.740 nm) requires
a mesh of $11\times11\times1$. The calculation precision (determined
with these test systems) is maintained throughout our investigation,
but the $k$-point mesh itself is not kept constant. For example,
moving from a $1\times1\times3$ to a $1\times1\times6$ supercell
in case of the (5,5) CNT allows us to employ a $1\times1\times5$
instead of a $1\times1\times10$ mesh while keeping the same precision.
Details regarding the testing of the Monkhorst-Pack mesh and energy
cut off are in the supplementary material, Tables S7-S11.}

\textcolor{black}{The binding e}nergies, $\Delta E$, for all systems
are 
\begin{equation}
\Delta E=E_{S}-E_{X}-E_{M},\label{eq:BindingEnergy}
\end{equation}
where $E_{S}$ is the total energy of the simulated system, $E_{X}$
is the energy of the adsorbate free substrate (graphene, CNT, or BNNT),
and $E_{M}$ is the energy of the isolated metal atom obtained at
the gamma point in a cell identical or very close (in order to break
symmetry) to the one used for $E_{S}$. Thus, for example, if $E_{S}$
is obtained from a $3\times3$ graphene super cell, so is $E_{M}$.

We further employ a tight-binding (TB) Hamiltonian for a system consisting
of a lattice of $n$ atoms, which represents a model for the nanotube/graphene
sheet, where each atom is a two-level system with degenerate orbitals.
The energy of each atomic level is set to zero. The atomic orbitals
of adjacent atoms overlap, introducing an electron hopping probability
between them with coupling energy $t$. A single atom, represented
by a single orbital, of zero energy $(\varepsilon_{M}=0)$, couples
to the first atom with strength $\beta$. The resulting Hamiltonian
$\hat{H}_{n}$ is a $2n+1$ square matrix of the form 
\begin{align}
\hat{H}_{n}= & [0]\oplus(A_{n}\otimes T)+M_{2n+1},
\end{align}
where $T$ is the orbital coupling matrix 
\begin{align}
T= & t\,\begin{bmatrix}1 & 1\\
1 & 1
\end{bmatrix},
\end{align}
$A_{n}$ is the adjacency matrix with $\{i,j\}$ element 
\begin{align}
[A_{n}]_{i,j}= & \delta_{i,j+1}+\delta_{i+1,j}+\delta_{1,i}\delta_{n,j}+\delta_{n,i}\delta_{1,j}\:,
\end{align}
and $M_{2n+1}$ is the matrix for the metal-nanotube/graphene interaction
\begin{align}
[M_{2n+1}]_{i,j}= & \beta(\delta_{1,i}\delta_{j,2}+\delta_{1,i}\delta_{j,3}+\delta_{2,i}\delta_{j,1}+\delta_{3,i}\delta_{j,1}).
\end{align}
Molecular orbital energies $\varepsilon_{k}$ are a result of diagonalizing
$\hat{H}_{n}$. By setting $\beta=0$, we recover the free systems
(i.e., isolated impurities and pristine substrates). The binding energy
is therefore the difference between the electronic energies of the
bonded and unbonded system. Ordering the set of eigenenergies $\{\varepsilon_{k}\}$
such that \linebreak{}
 $\varepsilon_{k}\leq\varepsilon_{k+1}$, the binding energy is $\sum_{k=1}^{n+1}\varepsilon_{k}(\beta)-\sum_{k=1}^{n+1}\varepsilon_{k}(\beta=0)$,
where the summation includes $n$ electrons for the substrate (half-filling)
plus one electron from the adatom \footnote{Disclaimer: Certain commercial products are identified in this paper
in order to specify the theoretical procedure adequately. Such identification
is not intended to imply recommendation or endorsement by the National
Institute of Standards and Technology nor is it intended to imply
that the software identified is necessarily the best available for
the type of work.}.

\noindent \textbf{\textsc{Results and Discussion}}

\textit{Adsorption Sites.} We study the adsorption of Al, Ti, Fe,
Ni, V, and Cu on graphene, CNTs, and BNNTs on several adsorption sites
to determine the most stable configuration. In the case of graphene,
these are the hollow (\textit{H}), top (\textit{T}), and bridge (\textit{B})
sites, representing adsorption in the center of a 6-fold carbon ring,
on top of a carbon, and between two carbon at\textcolor{black}{oms,
respectively. For adsorption on CNTs and BNNTs, we study the }\textit{\textcolor{black}{H}}\textcolor{black}{{}
site and two different }\textit{\textcolor{black}{B}}\textcolor{black}{{}
sites}\textit{\textcolor{black}{{} }}\textit{\textcolor{black}{\emph{(}}}\textit{\textcolor{black}{B1}}\textcolor{black}{{}
and }\textit{\textcolor{black}{B2}}\textcolor{black}{, }\textit{\textcolor{black}{BN1}}\textcolor{black}{{}
and }\textit{\textcolor{black}{BN2}}\textcolor{black}{{} in case
of BNNTs)}\textit{\textcolor{black}{\emph{.}}}\textit{\textcolor{black}{{}
}}\textcolor{black}{These represent bridge sites running parallel
to the growth direction }\textit{\textcolor{black}{(B1 / BN1)}}\textcolor{black}{{}
and a neighboring bridge site }\textit{\textcolor{black}{(B2 / BN2)}}\textcolor{black}{{}
at an angle of 60${^{\circ}}$ to the former with respect to the (8,0)
CNT. For the (5,5) CNTs (BNNTs) the }\textit{\textcolor{black}{B1}}\textcolor{black}{{}
(}\textit{\textcolor{black}{BN1}}\textcolor{black}{) site represents
those running perpendicular to the growth direction of the NT with
the }\textit{\textcolor{black}{B2}}\textcolor{black}{{} (}\textit{\textcolor{black}{BN}}\textcolor{black}{2)
site neighboring the former at an angle of 60${^{\circ}}$.{} Our
calculations show that, irrespective of the system size, the preferred
adsorption site remains the same. However, we do note small changes
in the metal-carbon or metal-boron/nitrogen bond length upon an increase
in system size. The adsorption sites are presented in Table S2 (in
the supplementary material) and compare well with those of earlier
investigations \citep{pasti_atomic_2018,manade_transition_2015,rohmann_interaction_2016}. }

\textit{General Considerations.} In the followi\textcolor{black}{ng,
we show and discuss the changes in binding energy upon an increase
in system size, as well as the effect of the isolated atom and inter-layer
spacing (vacuum gap). In this regard, the energy of the free atom
$E_{M}$, which might appear trivial, can have a significant impact
on the binding strength. While using the identical cell size for the
isolated atom and the adsorbed atom is a standard protocol, we examine
its effect since it is often unclear whether this protocol is employed.
In many calculations, the adsorbate, a single metal atom in our case,
is simply placed in a }\linebreak{}
 1 nm $\times$ 1 nm $\times$ 1 nm cell, or a small variation of
this to break the cubic symme\textcolor{black}{try. Not breaking the
cubic symmetry of the cell can lead to an incorrect estimation of
the ground state, as observed for Ni in our case. In our calculations,
placing the metal atoms in the respective cells from the actual adsorption
study leads to errors in the ground state of several tenths of an
electron volt for Al, Ti, and V. In addition to this, one also needs
to consider the lateral separation of adatoms in the $E_{X}$ term
in Eq.~\eqref{eq:BindingEnergy}. Employing a $4\times4$ supercell
of graphene (roughly, 1 nm $\times$ 1 nm) with a 1 nm and 2 nm vacuum
layer separation yields a difference in energy of 0.4 eV for Fe adsorbing
on the }\textit{\textcolor{black}{T}}\textcolor{black}{{} site, whereas
all other metals show a difference of less than 0.05 eV. This example
illustrates the effect of the layer spacing on the binding energy.
While adsorption at the }\textcolor{black}{\emph{T}}\textcolor{black}{{}
site results in a difference of 0.4 eV, the adsorption at the }\textcolor{black}{\emph{H}}\textcolor{black}{{}
site (which is still the most stable site) results in a difference
of less than 0.05 eV when going from a 1 nm to a 2 nm layer spacing.
The }change in energy of the pure, adsorbate free graphene sheets
is less than 0.03 eV in case of a \linebreak{}
 1 nm sheet separation, which is in agreement with earlier results
\citep{manade_transition_2015}.

However, to prevent the interaction with the periodic image, and to
obtain a fully converged binding energy, one needs to consider the
adsorbate covered system to determine the required cell size, since
(i) the adsorbate by itself requires a certain cell size and (ii)
the structural and electronic properties of the system will change
due to the introduction of the adatom, which can also require the
use of larger system cells. In some studies, a finite adsorbate coverage
might be of interest. However, to study the adsorption of a single,
isolated adsorbate, the system needs to be large enough to prevent
the interaction of periodic images, regardless of whether the added
atom is adsorbed to the surface or free.

\begin{figure*}
\centering
\includegraphics[width=2.45\columnwidth]{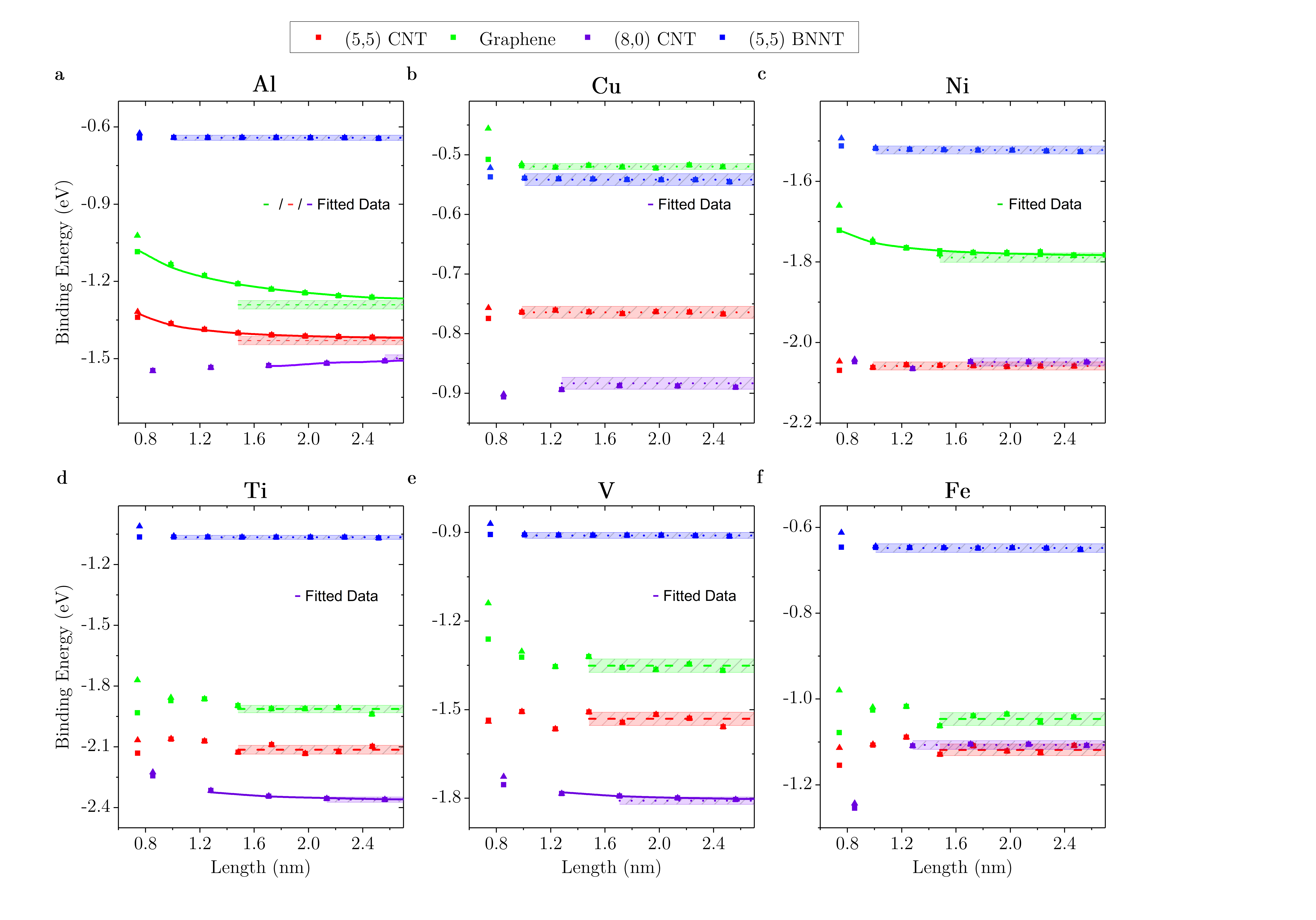}
\caption{\textbf{Binding energies versus system size.} Here, the blue, purple,
green, and red data points represent the insulating (5,5) BNNT, semiconducting
(8,0) CNT, zero-point semiconductor graphene, and the metallic (5,5)
CNT, respectively. The squares show the results using the absolute
ground state energy of the free atoms, whereas the triangles represent
the results from the free atom in the respective adsorption size cells.
A dotted line indicates fully converged DFT data, whereas dashed lines
indicate the extrapolated binding energy (numerical fitting to Eq.~8
in case of Al for the (5,5) CNT and graphene) or averaged binding
energy over the indicated data points (V, Fe, and Ti for graphene
and the (5,5) CNT). In all cases, the shaded area represents the error
in the binding energy originating from the DFT calculations. The solid
line represents a fit to the scaling form in Eq.~(\ref{eq:Scaling})
from the TB model. In the determination of the fitting parameters,
only data points for the longer NT system sizes and larger graphene
sheets are considered since Eq.~(\ref{eq:Scaling}) is the asymptotic
expression for the binding energy, see supplementary material. We
estimate the error variance $\delta E=\pm \sqrt{\sigma_{DFT}^{2}+\sigma_{\Delta E_{\infty}}^{2}}$
using the fitting error $\sigma_{\triangle E_{\infty}}$ Table S6 in the see supplementary material for more information). The estimated error in the DFT calculations
 $\sigma_{DFT}=0.01$~eV. In cases that use an average value for the
binding energy, the error represents the maximal difference in energy
of the data points considered for averaging. \label{fig:BindingEnergy}}
\end{figure*}

The effect of the free atom (essentially, direct interactions) is
clearly seen in Fig.~\ref{fig:BindingEnergy}, which gives the binding
energy versus the NT length or graphene sheet size. Here, the different
systems are represented by different colors, whereas the binding energy
is shown as squares and triangles. The squares represent the binding
ener\textcolor{black}{gy from the absolute ground state of the free
atom (the free atom in a cell large enough to prevent interactions
with its periodic image) and the triangles represent results from
the free atom in the respective adsorption cell. The latter is at
times very small and thus results in significant interactions of the
atom with its periodic image. In case of graphene, this is a stronger
effect due to the fact that it is a 2D material, while the NTs are
1D. Thus, the periodic image is `felt' from the $x$ and $y$ direction
in graphene while only in the $z$ direction in case of the NTs, since
one can extend the vacuum gap in the other directions. To provide
an additional perspective on these findings, we include a substrate-based
version of Fig. 2 in the supplementary material (Fig. S4).}

\textit{System Size and Metal Type.} Examining the trend of binding
energy versus system size, Fig.~\ref{fig:BindingEnergy}, we note
the following: (i) that insulating systems are least affected by the
system size, whereas metallic systems -- the (5,5) CNT and graphene,
which is turned into a metallic system after metal atom adsorption
-- exhibit the largest changes in $\Delta E$. (ii) We do not achieve
full convergence -- convergence to within the estimated DFT error
of 0.01 eV \footnote{We estimate the error in our DFT calculations by conducting multiple
test calculations with Al adsorbing to graphene. A variety of different
initial positions for Al are tested. The error results from the largest
difference in binding strength of all structures where Al is found
at the \textit{H} site (the preferred adsorptions site) after optimization.
The largest difference is noted as 0.006 eV. Therefore we estimate
an error of 0.01 eV for all metals and systems investigated.} -- of the binding energy for Ti, V, and Fe adsorbing to metallic
substrates for system sizes with linear dimension of greater than
2.4 nm. (iii) \textcolor{black}{For adsorption on the semiconducting
(8,0) CNT, there is a marginal but steady increase in the binding
energy for Al and a decrease for Ti and V }up to roughly\linebreak{}
 2.7 nm. Further increasing the length of the system shows that Al,
Ti and V are actually converged at a 2.7 nm length of the (8,0) CNT
(not shown in Fig.~\ref{fig:BindingEnergy}). The observations (i)
and (ii), together with the fact that free atom calculations are fully
converged for the larger cells (see the overlap of square and triangle
data points), indicates that there is a quasi-long range interaction
between metal adsorbates and their periodic images. This interaction
that proceeds through the substrate is strongest through metallic
substrates. We can thus distinguish between direct -- i.e., through
free space -- and indirect -- i.e., through the substrate -- mechanisms
for adsorbate-adsorbate interactions \ref{fig:Schematic}.

We note that Cu is an exception to observation (i), with a binding
energy that changes only as the system size increases from $\approx0.8$
nm to $\approx1$ nm for the \linebreak{}
 (5,5) CNT and graphene. A possible explanation could stem from the
fact that Cu is the only d$^{10}$ metal in our study and has, together
with Ni, the highest electronegativity among the metals we examine.
Considering Ni in this respect, it is not surprising that it shows
only small changes in $\Delta E$ for the metallic systems. Ni is
a d$^{9}$ metal, unlike Cu. However, upon adsorption Ni's 4s electron
is \textcolor{black}{filling} the d-orbitals, while a small portion
of its electron density is donated to graphene due to the higher electronegativity
of C compared to Ni. Ni's d-orbitals are thus close to, but less than,
full occupancy. One can therefore assume a similar behavior with Cu.
Indeed, Ni and Cu show an almost identical behavior for adsorption
to the insulating (5,5) BNNT and semiconducting (8,0) CNT, and are
essentially converged above about 1.5 nm for the metallic systems.

In contrast, Fe, Ti, and V, see Fig.~\ref{fig:BindingEnergy}d-f,
show larger, although still small, fluctuations in binding energy
even at the largest system sizes for the metallic systems.\textcolor{black}{{}
Further increasing the cell dimension to 3 nm, reducing the }$k$\textcolor{black}{-point
spacing by a factor of three, employing a gamma point centered grid
or eliminating the small aliasing errors in VASP calculations (via
sufficiently high settings) all fail to reduce or eliminate these
fluctuations and trends. It is likely that these fluctuations represent
physical, Friedel-like oscillations, as they appear upon binding to
metals and semiconductors. Ti, Fe, and V all have an open 3d shell,
whereas Cu and adsorbed Ni are d$^{10}$ and nearly d$^{10}$, respectively,
which may be the underlying cause of their different behavior. }

We further note the existence of many local minima
close to the identified global minimum for large cells of graphene
with the adsorption of Cu, Ni, and Fe. In order to identify the most
stable adsorption site, a variety of different geometries were tested
around where we believe the global minima is situated. This could cause oscillations in binding energies as well. However, upon extensive testing, it seems unlikely to be responsible for the
oscillations seen for Fe, but it may be causing the minute deviation
of the last data points for Ni and Cu in case of graphene and the
(5,5) CNT respectively
. 
\begin{figure}[t]
\centering
\includegraphics[scale=0.35]{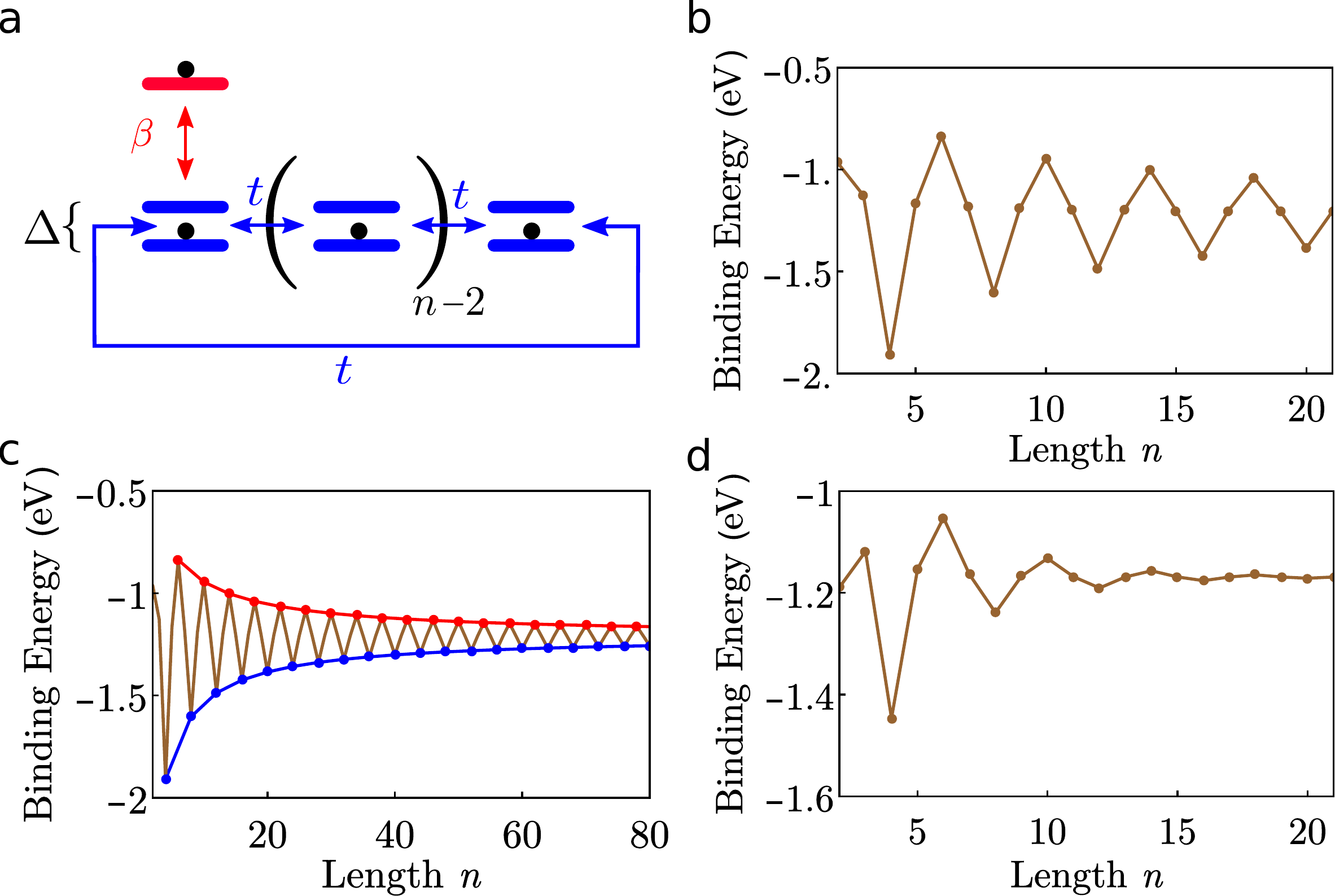}
\caption{\textbf{Tight-binding model for metal-nanotube/graphene
binding energies.} (a) A model one-dimensional
substrate consisting of a periodic lattice of $n$ two-level atoms,
a nearest neighbor interaction $t$, and adsorbate-substrate interaction
$\beta$ to the first atom in the system only. (b) Binding energies
versus the system length $n$ when the energy gap is zero, $\beta=1.2$
eV, and $t=1.25$ eV. (c) Binding energies for the same parameters as
in (b) but for the system length taking only values $n=4k+6$ (red)
and $n=4k+4$ (blue) with $k$ an integer. (d) Binding energies versus
$n$ with an energy gap $\triangle$ between the two levels of
0.4 eV (but otherwise the parameters are the same as in (b)), thus
giving a representation of an insulating or semiconducting system
$n$. \label{fig:TightBinding}} 
\end{figure}

To investigate these fluctuations further, we
examine the use of D3 (Grimme) calculations on metal adsorption to
the (5,5) CNT. Although the absolute binding energies are different,
the overall trend and fluctuations remain the same, see Fig. S6 in
the supplementary material for more details.

\textcolor{black}{Next, we examine the tight-binding model of Fig.~\ref{fig:TightBinding}a.
This particular model has no direct adsorbate-adsorbate interaction
and thus allows us to examine solely the effect of indirect interactions.
The size of the system is proportional to the number $n$ of atoms
in the system. Figure \ref{fig:TightBinding}b,c shows that the binding
energies converge for large system sizes, following an oscillatory
pattern. Convergence is faster when there is an energy gap, as seen
in Fig.~\ref{fig:TightBinding}d, where it is also seen that the
oscillations are more effectively damped. The latter behavior is qualitatively
similar to much of t}he oscillatory behavior seen in the DFT calculations.
Moreover, how the system size is increased can lead to a monotonic
increase ($n=4k+4$ with $k$ an integer) or decrease ($n=4k+6$)
towards the asymptotic (isolated impurity) limit. Thus, the range
of behavior given by DFT indicates intrinsic, physical behavior and
not something particular to the method of calculation.

For the tight-binding model in Fig.~\ref{fig:TightBinding}a, we
can provide an analytic estimate for the convergence with the cell
length. From the exact energy eigenvalues and eigenstates for the
unbound system and taking the coupling strength $\beta$ as the perturbation
parameter, the binding energy versus $n$ is 
\begin{equation}
\Delta E=\left(\frac{\beta^{2}}{t}-\frac{\beta^{4}}{8t^{3}}\right)\times I_{n},\label{eq:EnergyTB}
\end{equation}
up to 4$^{{\rm th}}$ order in \textbf{$\beta$}. In Eq.~\eqref{eq:EnergyTB},
the term $I_{n}$ carries the length dependence and is explicitly
given by 
\begin{equation}
I_{n}=\frac{1}{n}\sum_{k\geq n/4}^{3n/4}\cos\left(2\pi k/n\right),
\end{equation}
where we use the fact that the Fermi level is at half filling \footnote{Note that the summation index $k$ takes integer values between $\frac{n}{4}$
and $\frac{3n}{4}$.}. The quantity $I_{n}$ has the form of a Riemann sum, and converges
to $-1/\pi$, as $n$ goes to infinity. As a result, $\Delta E$ reaches
the asymptotic value as some power of $1/n$ \citep{chui_concerning_1971}.
Since the linear dimension $L$ of the system is proportional to $n$,
the binding energy $\Delta E$ decays as 
\begin{equation}
\Delta E(L)=\Delta E(\infty)+\frac{B}{L^{2}},\hspace{0.5cm}L\rightarrow\infty,\label{eq:Scaling}
\end{equation}
\textcolor{black}{where }$B$\textcolor{black}{{} is a coefficient
that depends on the specific function in the Riemann sum (for} more
details see the supplemental material). This coefficient can be both
positive or negative (see Fig.~\ref{fig:TightBinding}\textcolor{black}{c)
and can also oscillate (but remain bounded so that the asymptotic
decay is $1/L^{2}$ but oscillatory). We emphasize that Eq.}~\textcolor{black}{(\ref{eq:Scaling})
describes the asymptotic decay to the limit binding energy }$\Delta E(\infty)$\textcolor{black}{{}
and the exact dependence on $L$ for smaller cells might be different. }

\textcolor{black}{This decay and other tight-binding results are in
qualitative agreement with the DFT results. Both show that (i) a larger
sheet or longer NT is required to obtain a converged binding energy
in case of a metallic system, see the respective graphene and (5,5)
CNT (green and red data points, respectively) systems of Al, Cu, and
Ni in Fig.}~\ref{fig:BindingEnergy}\textcolor{black}{{} a-c in
comparison to Fig.~}\ref{fig:TightBinding}\textcolor{black}{c. (ii)
Systems with a band gap converge faster than their metallic counterparts,
see the (8,0) CNT (purple data points) and }\linebreak{}
 \textcolor{black}{(5,5) BNNT (blue data points) semiconducting and
insulating systems versus the metallic graphene and (5,5) CNT systems
(green and red data points, respectively) in Fig.}~\ref{fig:BindingEnergy}\textcolor{black}{{}
in comparison to Fig.}~\ref{fig:TightBinding}\textcolor{black}{b
versus 3d. (iii) The tight-binding model shows fluctuations in the
binding energy for metallic systems, see Fig.}~\ref{fig:TightBinding}\textcolor{black}{b
and c, even at a larger system size which we also note in case of
graphene and the (5,5) CNT for the open shell systems of Fe, Ti, and
V, see Fig.}~\ref{fig:BindingEnergy}\textcolor{black}{d-f. Thus,
we provide evidence that the fluctuation noted for Ti, V, and Fe are
real and do not stem from particular choices of DFT parameters or
methods.}

In addition, the decay form, Eq\textcolor{black}{.~}\eqref{eq:Scaling},
gives a method to scale finite-size results and extrapolate to the
isolated impurity li\textcolor{black}{mit.} For example, Fig.~\ref{fig:BindingEnergy}
a shows Eq.~\eqref{eq:Scaling} fit to the binding energy decay for
Al on the (5,5) CNT and graphene (solid red and green lines respectively),
along with the value of $\Delta E(\infty)$ and error bars from the
respective DFT calculations. In principle, it is possible to employ
Eq.~\eqref{eq:Scaling} for all DFT results, including the oscillatory
behavior. However, the accessible range of length scales due to computational
resources and the estimated error in DFT make fitting oscillatory
$B$ error-prone. There are not enough data points, and even if there
were, the decay would put the value at finite $L$ to within the error
of ou\textcolor{black}{r DFT calculations (0.01 eV). We thus separate
the DFT results into three classes, ones with monotonic decay (Al
on the (5,5) CNT, graphene, and the (8,0) CNT, as well as Ti and V
on the (8,0) CNT), ones with rapid convergence {[}Cu and Ni; and all
metals on BNNTs{]}, and those with oscillatory behavior (Ti, V, and
Fe on the (5,5) CNTs and graphene). The first two classes can make
use of the scaling form, although the second class does not need it
since the result is effectively $\Delta E(\infty)$, except for the
smallest cell sizes considered. Error bars for the cases are defined
in terms of the fit errors. For the third class, we propose a heuristic
method: take a range of the largest cell sizes accessible (in our
case, cells from 1 nm or 1.5 nm to 2.7 nm), c}ompute the average $\Delta E$
within that range, which averages over some of the oscillations, and
use the standard deviation to quantify the error. The isolated impurity
binding energies for all metals and systems is shown in Table S3 (see
supplementary material).

\noindent \textbf{\textsc{Conclusion}}

\textcolor{black}{Our results yield the magnitude of direct and indirect
interactions for metal atom adsorption on common substrates (graphene
and metallic, semiconducting, and insulating NTs). The factors that
influence the strength of these interactions and/or the computed binding
strengths are (i) the nature of the substrate (metallic, semiconducting
or insulating), (ii) the closed or open shell structure of the metal,
(iii) the method of computing the isolated atom energies, and (iv)
the size of the simulation cell (both in the direction(s) of the substrate
and any vacuum gap between structures). Insulating systems generally
require the smallest simulation sizes followed by semi-conducting
and metallic systems. Open shell systems, such as Fe, V, and Ti, do
not fully converge, but rather show a fluctuating -- but decaying
-- binding energy at larger simulation cell sizes. We derive a scaling
relation that permits the extrapolation of smaller simulation cell
results out to the infinite simulation cell limit (i.e., a fully isolated
impurity). This enables a more effective calculation of binding in
slowly converging systems, such as Al on graphene and the (5,5) CNT.
Overall, this scaling relation and a simple heuristic for oscillatory
systems yields a framework for determining binding energies and providing
error estimates.}

\noindent \textbf{\textsc{Supplementary Material}}

See supplementary material\textcolor{black}{{} for the convergence
tests, the D3 versus D2 results, a detailed derivation of Eq.~(8),
the binding energies from 1D-tight binding models, and the criteria
for the numerical fittings.}

\noindent \textbf{\textsc{\textcolor{black}{Acknowledgments}}}

We thank J. Elenewski and S. Sahu for helpful discussions. C. R. and
M. A. O. acknowledge the support under the Cooperative Research Agreement
between the University of Maryland and the National Institute of Standards
and Technology Physical Measurements Laboratory, Award 70NANB14H209,
through the University of Maryland.  
\bibliography{Single_atoms_on_CNTs_and_BNNTs_abbreviations}
 

\end{document}


\global\long\def\avg#1{\langle#1\rangle}

\global\long\def\p{\prime}

\global\long\def\dg{\dagger}

\global\long\def\ket#1{|#1\rangle}

\global\long\def\bra#1{\langle#1|}

\global\long\def\proj#1#2{|#1\rangle\langle#2|}

\global\long\def\inner#1#2{\langle#1|#2\rangle}

\global\long\def\tr{\mathrm{tr}}

\global\long\def\pd#1#2{\frac{\partial#1}{\partial#2}}

\global\long\def\spd#1#2{\frac{\partial^{2}#1}{\partial#2^{2}}}

\global\long\def\der#1#2{\frac{d#1}{d#2}}

\global\long\def\im{\imath}

\global\long\def\onlinecite#1{\cite{#1}}

\title{Supplementary Material -- Metal adsorbate interactions and the convergence of density functional
calculations}

\author{Christoph Rohmann}

\address{Biophysics Group, Microsystems and Nanotechnology Division, Physical
Measurement Laboratory, National Institute of Standards and Technology,
Gaithersburg, MD 20899, USA}

\address{Maryland NanoCenter, University of Maryland, College Park, Maryland
20783, United States}

\author{Maicol A. Ochoa}

\address{Biophysics Group, Microsystems and Nanotechnology Division, Physical
Measurement Laboratory, National Institute of Standards and Technology,
Gaithersburg, MD 20899, USA}

\address{Maryland NanoCenter, University of Maryland, College Park, Maryland
20783, United States}

\author{Michael Zwolak}

\address{Biophysics Group, Microsystems and Nanotechnology Division, Physical
Measurement Laboratory, National Institute of Standards and Technology,
Gaithersburg, MD 20899, USA}

\date{\today}

\maketitle

\section{Single-atom binding}

\vspace{-0.6cm}

\begin{center}
\begin{table}[H]
\centering{}%
\begin{tabular}{|c|c|c|c|c|c|c|c|c|}
\hline 
CNT  & \multicolumn{8}{c|}{Length in \textit{c} (nm)}\tabularnewline
\hline 
(8,0)  & 0.854  & 1.281  & 1.708  & 2.135  & 2.562  & \multicolumn{2}{c}{} & \tabularnewline
\hline 
(5,5)  & 0.741  & 0.988  & 1.235  & 1.482  & 1.729  & 1.976  & \multicolumn{1}{c|}{2.223} & 2.470\tabularnewline
\hline 
\hline 
BNNT  & \multicolumn{8}{c|}{Length in \textit{c} (nm)}\tabularnewline
\hline 
(5,5)  & 0.755  & 1.007  & 1.259  & 1.1511  & 1.763  & 2.014  & \multicolumn{1}{c|}{2.266} & 2.518\tabularnewline
\hline 
\hline 
Graphene & \multicolumn{8}{c|}{Dimensions in \textit{a} and \textit{b} (nm)}\tabularnewline
\hline 
 & 0.740  & 0.987  & 1.234  & 1.481  & 1.728  & 1.974  & \multicolumn{1}{c|}{2.221} & 2.468\tabularnewline
\hline 
\end{tabular}
\vspace{-0.1cm}
\caption{\textbf{Cell parameters}. The dimensions of the super cells in direction
\textit{c} for the different nanotube lengths, as well as in \textit{a}
and \textit{b} for the width and length of the graphene sheet. \label{tab:UnitCells}}
%
\begin{tabular}{|c|c|c|c|c|c|c|}
\hline 
Metal/Systems  & Al  & Ti  & V & Fe  & Ni & Cu \tabularnewline
\hline 
\hline 
Graphene  & \textit{H}  & \textit{H}  & \textit{H}  & \textit{H}  & \textit{H} & \textit{T} \tabularnewline
\hline 
(5,5) CNT  & \textit{H}  & \textit{H}  & \textit{H}  & \textit{H}  & \textit{T} & \textit{B} \tabularnewline
\hline 
(8,0) CNT  & \textit{H}  & \textit{H}  & \textit{H}  & \textit{H}  & \textit{B} & \textit{B} \tabularnewline
\hline 
\hspace{0.24cm}(5,5) BNNT  & \textit{BN1}  & \textit{H}  & \textit{T} & \textit{BN1}  & \textit{T} & \textit{BN1} \tabularnewline
\hline 
\end{tabular}
\vspace{-0.1cm}
\caption{\textbf{Adsorption sites}. The preferred adsorption sites of the metals on the respective NTs and graphene with {\it H},{\it  T}, and {\it B} indicating the adsorption to the hollow, top, and bridge sites, respectively. The {\it BN1} site represents a bridging position between N and B running perpendicular to the growth direction of the BNNT. The {\it T} site adsorption in the case of Ni and V adsorbing on the (5,5) BNNTs takes place above a N atom. For each system, three different adsorption sites were tested. In each case, the preferred adsorption site remained the same regardless of system size. For example, although the {\it H}, {\it T}, and {\it B} are always tested, the {\it H} site was found to be the most stable adsorption site for Al adsorbing on graphene regardless of the system size. \label{tab:AdsorptionSites}}
\end{table}
\end{center}

\vspace{-1.6cm}

\begin{center}
  \begin{table}[H]
  \begin{center}
  \begin{tabular}{|c|c|c|c|c|c|c|} \hline
    {\bf $\Delta E$ (eV)}& Al & Ti & V\\ \hline \hline
    Graphene& -1.29 $\pm$ 0.016&-1.91 $\pm$ 0.018&-1.35 $\pm$ 0.023\\ \hline
    (5,5) CNT&-1.43 $\pm$ 0.016&-2.11 $\pm$ 0.022&-1.53 $\pm$ 0.022\\ \hline
    (8,0) CNT&-1.51 $\pm$ 0.013&-2.38 $\pm$ 0.013&-1.79 $\pm$ 0.012\\ \hline
    \hspace{0.24cm}(5,5) BNNT&-0.64 $\pm$ 0.010&-1.07 $\pm$ 0.010&-0.91 $\pm$ 0.010\\ \hline  \hline 
    {\bf $\Delta E$ (eV)} & Fe & Ni & Cu\\ \hline \hline
    Graphene &-1.05 $\pm$ 0.015&-1.78 $\pm$ 0.012&-0.52 $\pm$ 0.010 \\ \hline
    (5,5) CNT&-1.12 $\pm$ 0.013&-2.06 $\pm$ 0.010&-0.76 $\pm$ 0.010 \\ \hline
    (8,0) CNT&-1.11 $\pm$ 0.010&-2.05 $\pm$ 0.010&-0.89 $\pm$ 0.010 \\ \hline
    \hspace{0.24cm}(5,5) BNNT&-0.91 $\pm $0.010&-1.52 $\pm$ 0.010&-0.54 $\pm$ 0.010 \\ \hline 
  \end{tabular}
  \end{center}
\vspace{-0.1cm}
\caption{{\bf Binding Energies}. $\Delta E_X(\infty)$ for $X$ = Al, Ti, V, Fe, Ni, and Cu on different substrates (Graphene, CNT, BNNT). These values are obtained from fully converged DFT calculations by either fitting the data to the form in Eq.~(8),  or by averaging over the energies for the largest systems employed. See Fig.~2 and the main text for the error calculation method and additional details.}
  \end{table}
\end{center}


\section{1D-tight binding model}
\vspace{-0.3cm}
In this section, we calculate the binding energy for the tight-binding model in Fig.~3, reported within the main text in Eq.~(6). In summary, starting from the eigenvalues and eigenvectors for the periodic 1D-lattice, we calculate the energy for metal-nanotube system utilizing perturbation theory, where the metal-nanotube coupling strength $\beta$ is taken as the perturbation parameter. 

First, we consider a simple model consisting of $n$ single levels, tunneling constant $t$, and  Hamiltonian
\begin{align}\tag{S1}
  [ \hat H_n ]_{i,j} = t (\delta_{i,j+1}+\delta_{i+1,j}+\delta_{i,1}\delta_{j,n}+\delta_{i,n}\delta_{j,1} ).\label{eq:H1}
\end{align}
The system is periodic and consequently  the Hamiltonian is invariant under discrete space translations. This also implies that each eigenvector $| \alpha_l \rangle$ -- that we find by solving the eigenvalue equation $\hat H_n | \alpha_l \rangle = \alpha_l | \alpha_l \rangle $ for $l = 1, \dots, n$ -- is periodic. Thus, the unnormalized complex eigenvector is  
\begin{align}\tag{S2}
  | \alpha_l \rangle  = ( e^{i \frac{2 \pi}{n}l}, e^{i \frac{4 \pi}{n}l}, \dots, e^{i \frac{2 \pi k}{n}l},\dots, e^{i \frac{2 \pi (n-1)}{n}l}, 1)^T,  
\end{align}
and from the eigenvalue equation we obtain the relation
\begin{align}\tag{S3}
  t e^{i \frac{2 \pi (k-1)}{n}l} - \alpha_{l} e^{i \frac{2 \pi k }{n}l} + t e^{i \frac{2 \pi (k+1)}{n}l} = 0
\end{align}
from which we find the $l$ eigenenergy $\alpha_l$
\begin{align}\tag{S4}
  \alpha_l ={}&2 t \cos\left(\frac{2 \pi l}{n}\right).
\end{align}
We notice that the complex conjugate of $|\alpha_l \rangle $ is also an eigenvector with eigenvalue $\alpha_l$, and consequently $(1/2)(|\alpha_l \rangle +|\alpha_l \rangle^*)$ is the real eigenvector. The normalized real $l$ eigenvector is
\begin{align}\tag{S5}
  | \alpha_l \rangle = \sqrt{\frac{2}{n}} \left( \cos\left(\frac{2 \pi}{n}l\right), \dots, \cos\left(\frac{2 \pi k}{n}l\right),\dots, 1\right)^T.
\end{align}
Finally, the energy for this system, with electron density $n_e / n$, is the sum of the lowest $n_e$ eigenvalues.

Next, we consider that an atom, represented by a single enegy level of zero energy, binds to the first site in the periodic lattice, with coupling strength $\beta$. The Hamiltonian for the composite system is
\begin{align}\tag{S6}
  \hat H ={}& [0]\oplus \hat H_n + \beta \hat V,
\end{align}
where $\hat V$ is an $(n+1)$-dimensional square matrix with entries $[\hat V]_{i,j} = \delta_{i,1}\delta_{j,2}+\delta_{i,2}\delta_{j,1}$. Taking $\beta$ as the perturbation parameter, the eigenvectors basis for the unperturbated system consists of 
\begin{align}\tag{S7}
  | \lambda_0^{(0)} \rangle ={}& (1,0, \dots, 0) \\
  | \lambda_l^{(0)} \rangle = {}& (0) \oplus | \alpha_l \rangle \hspace{0.8cm} ( 1 \le l \le n ),\tag{S8}
\end{align}
with zeroth order eigenenergies $\lambda_0^{(0)} =0$ and $\lambda_l^{(0)} = \alpha_l$. We can write the eigenenergies $\lambda_l$ for the composite system up to second order 
\begin{multline}\tag{S9}
  \lambda_l^{(2)}(\beta) = \lambda_l^{(0)}+ \beta \langle \lambda_l^{(0)} | \hat V| \lambda_l^{(0)} \rangle\\+ \beta^2 \sum_{r \neq l }\frac{\left|\langle \lambda_l^{(0)}| \hat V | \lambda_r^{(0)}\rangle \right|^2}{\lambda_l^{(0)}- \lambda_r^{(0)}},
\end{multline}
\begin{figure}[t]
  \centering
  \includegraphics[scale=0.55]{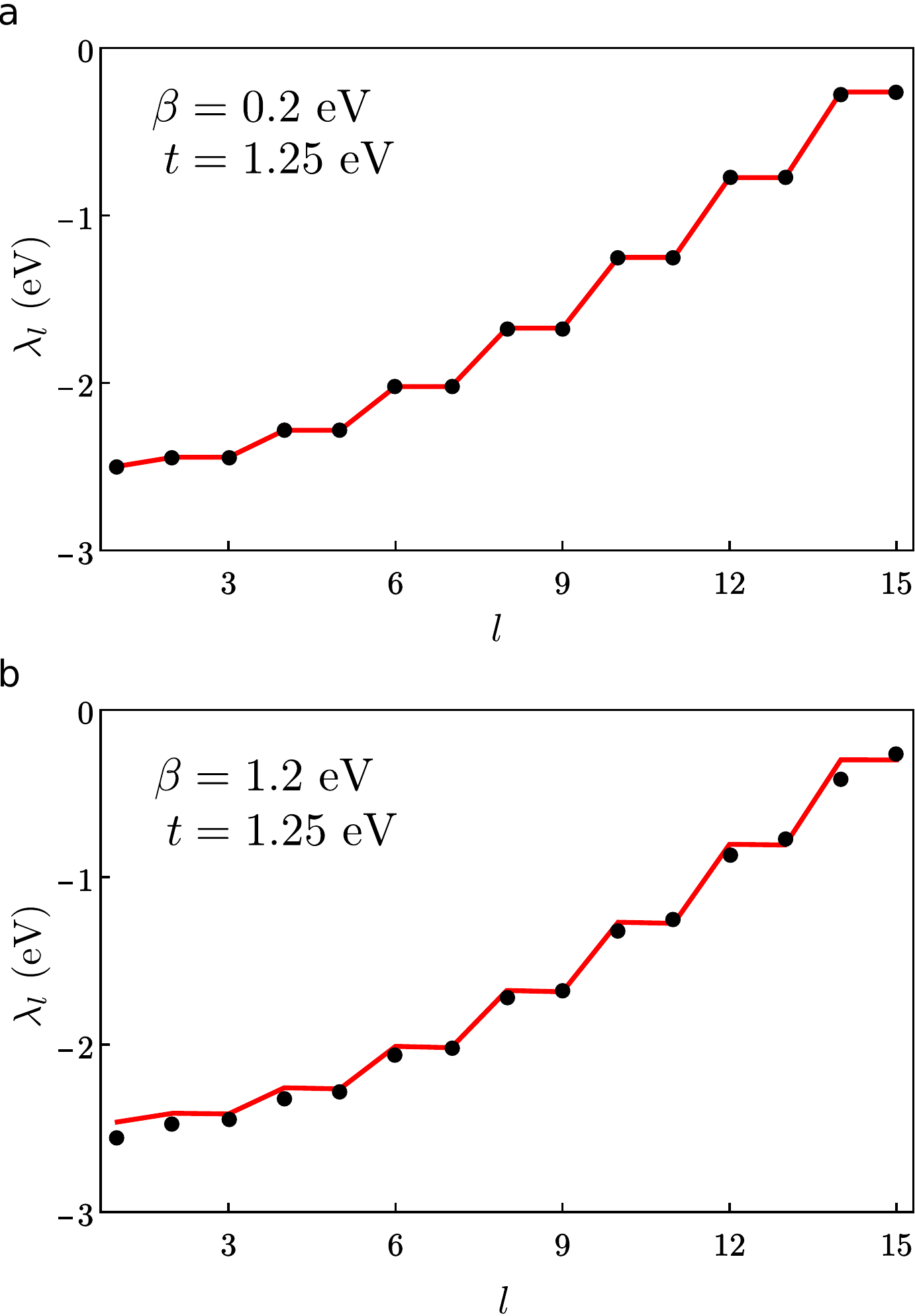}
  \caption{{\bf Exact and approximate eigenvalues of the tight-binding model}.  Comparison between the exact (dots, black) and approximate lowest eigenvalues (solid, red) for the 1D tight-binding model and unit size cell $n = 30$. Here $\lambda_l$ is the eigenvalue for state $l$. }
  \label{fig:energies}
\end{figure}

\noindent which is further simplified by noticing that
\begin{align}
 \langle \lambda_l^{(0)} | \hat V| \lambda_l^{(0)} \rangle ={}& 0 \tag{S10} \label{eq:S10}\\ 
  \langle \lambda_r^{(0)} | \hat V| \lambda_l^{(0)} \rangle ={}& 0 \hspace{2.3cm} (r,l \ge 1) \tag{S11}\\
   \langle \lambda_0^{(0)} | \hat V| \lambda_l^{(0)} \rangle ={}& \sqrt{\frac{2}{n}} \cos \left(\frac{2 \pi }{n} l \right) \hspace{0.5cm} (l \ge 1) \tag{S12}. \label{eq:S12}
\end{align}
Therefore
\begin{align}
     \lambda_0^{(2)}(\beta) ={}& \lambda_0^{(0)}= 0, \tag{S13}\\
   \lambda_l^{(2)}(\beta) ={}& \lambda_l^{(0)}+ \frac{\beta^2}{n t} \cos \left( \frac{2 \pi}{n} l \right) \hspace{0.5cm} (l\geq 1). \tag{S14}
\end{align}
If $n = 2 n_e$, then the binding energy up to second order is
\begin{align}
  \Delta E_n^{(2)} ={}& \sum_{l < {\rm LUMO}} \lambda_l^{(2)}(\beta) - \lambda_l^{(0)}  \tag{S15}\\
  ={}& \frac{\beta^2}{n t }\sum_{l \geq n/4 }^{l \leq 3 n/4} \cos\left( \frac{2 \pi}{n} l\right)\tag{S16}
\end{align}
The present analysis can be extended to include higher order corrections in the perturbation expansion. Since several matrix elements in the perturbation Hamiltonian cancel, Eqs.~\eqref{eq:S10}-\eqref{eq:S12}, we can carry out the full summation over higher order terms in the perturbation. For example, the forth order form of the binding energy is

\begin{align}
  \Delta E_n^{(4)} ={}& \left( \frac{\beta^2}{t}-\frac{\beta^4}{2 t^3 } \right) I_n \tag{S17},
\end{align}
where $I_n$ is defined in Eq. (6).

 These results can be extended to the model in Fig.~3, where each site has two degenerate levels ($\Delta =0$). For this, notice that the tunneling matrix $T$ in Eq.~(3) is singular, and therefore the rank of the Hamiltonian $\hat H_n$ in Eq.~(2) for $\beta=0$ is $n$. As a result, the system has $n$ zero eigenenergies and, after performing elementary operations, a submatrix of size $n$ of the form in Eq.~\eqref{eq:H1} is recovered with $t$ replaced by $2 t$.

Returning to the second order expression for the binding energy, the limit value, $\Delta E^{(2)} = \lim_{n \to \infty} \Delta E_n^{(2)}$, is the real binding energy for a single atom. We calculate this limit as
\begin{align}
  I = \lim_{n \to \infty} I_n ={}& \lim_{n \to \infty} \frac{1}{n}\sum_{l \geq n/4 }^{l \leq 3 n/4} \cos\left( \frac{2 \pi}{n} l\right) \tag{S18}\\
={}& \frac{1}{2 \pi} \lim_{n \to \infty} \sum_{l \geq n/4 }^{l \leq 3 n/4} \frac{2 \pi}{n} \cos\left( \frac{2 \pi}{n} l\right) \tag{S19}
\end{align}
Next, we let $\theta_l = (2 \pi/n) l$ and $\Delta \theta = \theta_{l+1} - \theta_l = (2 \pi/n)$, such that
\begin{align}
  I ={}& \frac{1}{2 \pi} \lim_{n \to \infty} \sum_{\theta_l \geq \pi/2 }^{3 \pi/2} \Delta \theta \cos\left( \theta_l\right) \tag{S20}\\
  ={}& \frac{1}{2 \pi} \int_{\pi/ 2}^{3\pi/2} d\theta \cos (\theta) = -\frac{1}{\pi} \tag{S21} \label{eq:In}.
\end{align}
Therefore, 
\begin{align}
  \Delta E^{(2)} ={}&  - \frac{\beta^2}{t \pi} \tag{S22}.
\end{align}
An alternative form for $I$ that we use in the next section, resulting from the change of variables $y = (2 / \pi) \theta$ is
\begin{align}
  I ={}& \frac{1}{4} \int_{1}^{3} d y \cos \left( \pi \frac{y}{2}\right) \tag{S23} .
\end{align}

\begin{figure}[t]
  \centering
  \includegraphics[scale=0.55]{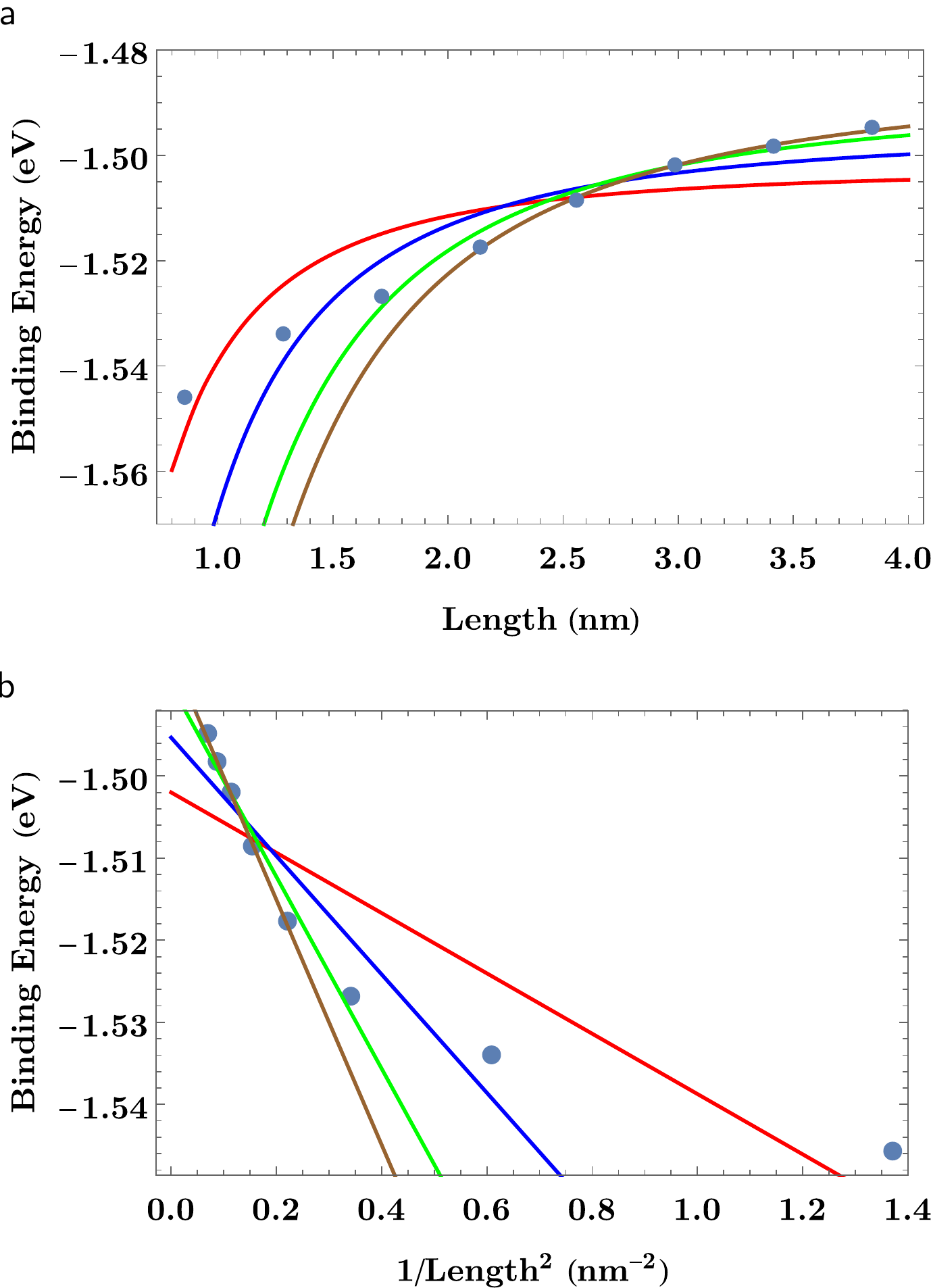}
  \caption{{\bf Binding Energy Dependence Al/(8,0) CNT}.  DFT Binding energies for Al in CNT (8,0) as a function of (a) unit cell size and (b) the inverse squared unit cell size. We include fitting curves Eq.~(8) for several fitting parameters $B$ and $\Delta E(\infty)$ obtained from the last $m$ points: (red) $m = 8$, (blue) $m = 7$, (green) $m = 6$, and (brown) $m = 5$. The value for the fitting paremeters are in reported in Table \ref{tab:Alfit}.    \label{fig:fitAl}}.
\end{figure}

\begin{figure}[t]
  \centering
  \includegraphics[scale=0.55]{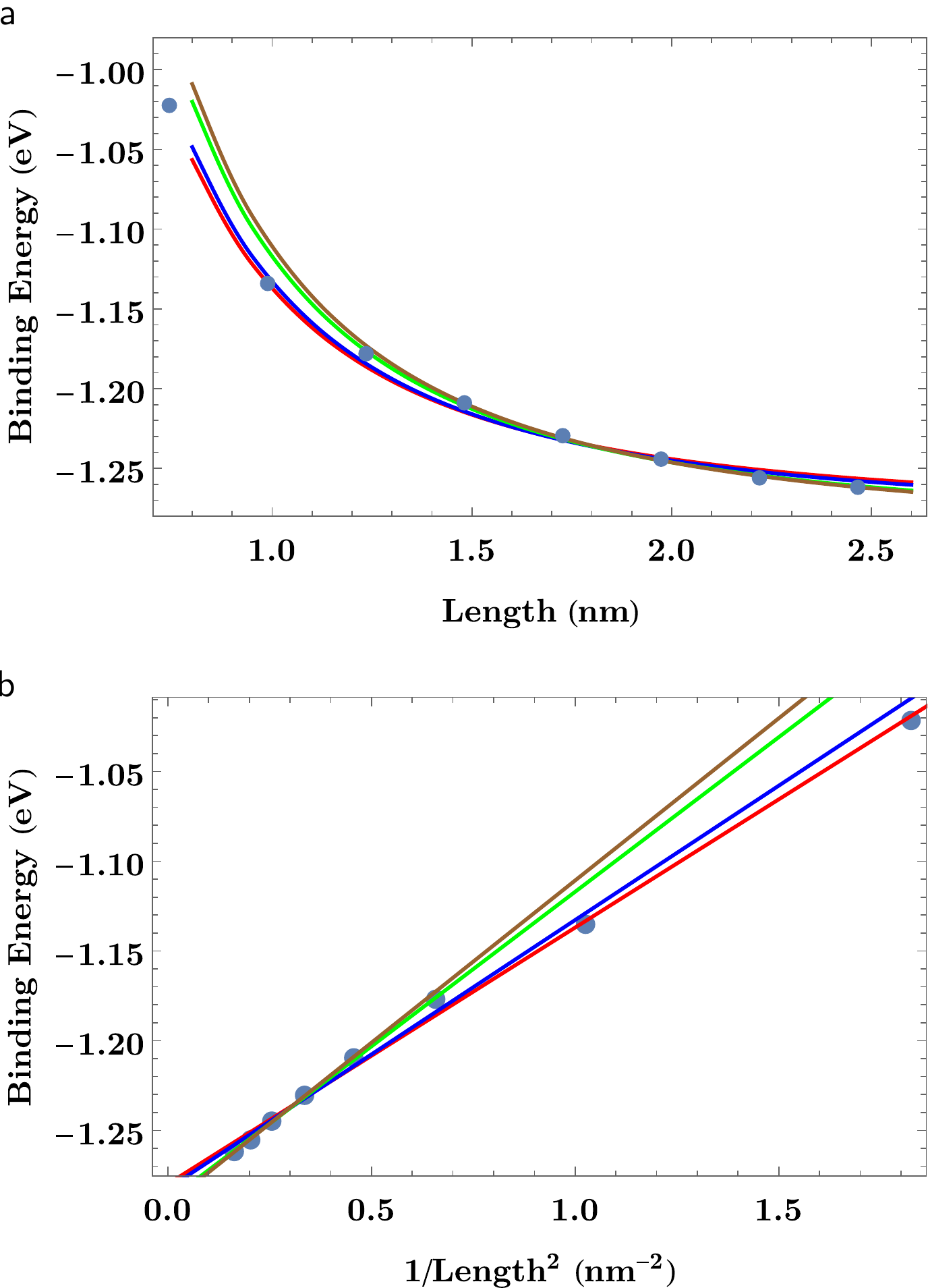}
  \caption{{\bf Binding Energy Dependence Al/Graphene}. DFT Binding energies for Al in Graphene as a function of (a) unit cell size and  (b) the inverse squared unit cell size. We include fitting curves Eq.~(8) for several fitting parameters $B$ and $\Delta E(\infty)$ obtained from the last $m$ points: (red) $m = 8$, (blue) $m = 7$, (green) $m = 6$, and (brown) $m = 5$. The value for the fitting paremeters are in reported in Table \ref{tab:Alfitgraphene}.    \label{fig:fitAlgraphene}}.
\end{figure}

\vspace{-0.8cm}
\section{Asymptotic decay}
\vspace{-0.2cm}
In this section we show that for the tight-binding model, binding energies converge to their asymptotic value as the inverse of the square of the length of the unit cell ($1/n^2$). This result motivates the fitting function Eq.~(8). The Riemman sum in Eq.~\eqref{eq:In} can be writen as an integral, after introducing the floor function $\lfloor x \rfloor$. First we note that
\begin{align}
  \Delta \theta \cos ( \Delta \theta l) ={}& \frac{2 \pi}{n} \int_l^{l+1} dx \cos \left(\frac{2 \pi}{n} \lfloor x \rfloor \right) \tag{S24}\\
  ={}& \frac{\pi}{2} \int_{4l/n}^{4(l+1)/n} dy \cos \left(\frac{2 \pi}{n} \left \lfloor \frac{n y}{4} \right \rfloor \right) \tag{S25}
\end{align}
\vspace{-0.2cm}

Therefore
\vspace{-0.3cm}
\begin{align}
  I_n  ={}& \frac{1}{4} \int_1^{3+4/n} dy \cos \left(\frac{2 \pi}{n} \left \lfloor \frac{n y}{4} \right \rfloor \right) \tag{S26} 
\intertext{and}
  I_n - I ={}& \frac{1}{4} \int_{1} ^{3} dy \left[  \cos \left (\frac{2 \pi }{n} \left \lfloor \frac{n y}{4} \right \rfloor \right) - \cos \left(\frac{2 \pi}{n} \frac{n y}{4} \right) \right]\notag\\
  &\hspace{0.5cm}+ \frac{1}{4} \int_3^{3 + 4/n} dy \cos \left(\frac{2 \pi}{n} \left \lfloor \frac{n y}{4} \right \rfloor \right) \tag{S27} \label{eq:Indiff}.
\end{align}

For $n = 4 m$, the integrand in the second term in Eq.~\eqref{eq:Indiff}  equals $\cos(3 \pi/ 2) = 0$. Next, we write $\cos \left(\frac{2 \pi}{n} \frac{n y}{4} \right)$ as a Taylor expansion up to second order around the integer $\left \lfloor \frac{n y}{4} \right \rfloor$
\begin{align}
  \cos \left(\frac{2 \pi}{n} \frac{n y}{4} \right) ={}& \cos \left(\frac{2 \pi}{n} \left \lfloor \frac{n y}{4}\right \rfloor \right) \notag\\
  & - \left( \frac{n y}{4} - \left \lfloor \frac{n y}{4}\right \rfloor \right) \left(\frac{2 \pi}{n}\right) \sin \left(\frac{2 \pi}{n} \left \lfloor \frac{n y}{4}\right \rfloor \right) \notag\\
  & + \frac{1}{2} \left( \frac{n y}{4} - \left \lfloor \frac{n y}{4}\right \rfloor \right)^2 \left( \frac{2 \pi}{n} \right)^2 \cos \left(\frac{2 \pi}{n} \left \lfloor \frac{n y}{4}\right \rfloor \right) \tag{S28}
\end{align}
\noindent and substitute into Eq.~\eqref{eq:Indiff}. We notice that $\sin \left(\frac{2 \pi}{n} \left \lfloor \frac{n y}{4}\right \rfloor \right)$ converges to the function $\sin \left(\frac{2 \pi}{n} \frac{n y}{4} \right)$, which is an odd function with respect to $y=2$ and therefore
\begin{align}
   \int_{1} ^{3}\sin \left(\frac{2 \pi}{n} \left \lfloor \frac{n y}{4}\right \rfloor \right) \to{}& 0, \tag{S29}
\end{align}

\noindent as $O(1/n)$. Consequently
\begin{align}
   I_n - I ={}& - \frac{\pi^2}{2 n^2} \int_{1} ^{3} dy \left( \frac{n y}{4} - \left \lfloor \frac{n y}{4}\right \rfloor \right)  \cos \left(\frac{2 \pi}{n} \left \lfloor \frac{n y}{4}\right \rfloor \right) \tag{S30}
\end{align}
which implies that
\begin{align}
  | I_n - I | \leq \frac{\pi^2}{n^2} . \tag{S31}
\end{align}

\section{Numerical Fittings}

\begin{center}
\begin{figure}[H]
  \centering
  \includegraphics[scale=0.57,trim=0 0 0 50, clip]{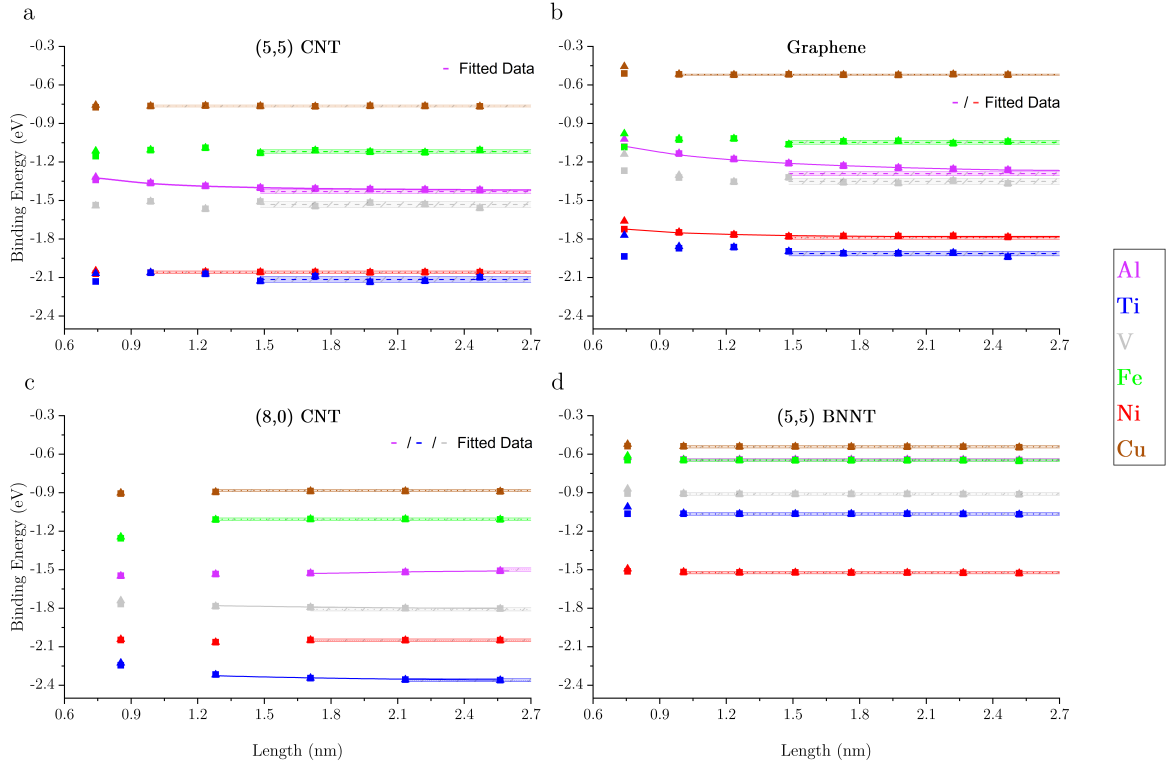}
  \vspace{-0.5cm}
  \caption{ {\bf Binding energies versus system size in a substrate-based comparison}. Here, the blue, red, grey, purple, green, and brown data points represent the add atoms Ti, Ni, V, Al, Fe and Cu, respectively. The squares show the results using the absolute ground state energy of the free atoms, whereas the triangles represent the results from the free atom in the respective adsorption size cells. A dotted line indicates fully converged DFT data, whereas dashed lines indicate the extrapolated binding energy (numerical fitting to Eq.~8 in case of Al for the (5,5) CNT and graphene) or averaged binding energy over the indicated data points (V, Fe, and Ti for graphene and the (5,5) CNT). In all cases, the shaded area represents the error in the binding energy originating from the DFT calculations. The solid line represents a fit to the scaling form in Eq. (8) from the TB model. Additional specifications are given in Fig.~2. \label{fig:fig2Sup}}
\end{figure}
\end{center}

\vspace{-0.3cm}
 In this section, we illustrate how the DFT binding energies are fitted to Eq.~(8) in the main text, and consequently how we obtain $\Delta E (\infty)$ in the case of binding energies that monotonically converge as a function of unit cell size. We consider Al in (8,0) CNT in Fig.~\ref{fig:fitAl} and Table~\ref{tab:Alfit}, and Al in graphene in Fig.~\ref{fig:fitAlgraphene} and Table~\ref{tab:Alfitgraphene} as representative examples. The result in Eq.~(8) is a good representation of the functional form of the binding energy for unit cell sizes that are large. This observation is confirmed in Figs.~\ref{fig:fitAl} and \ref{fig:fitAlgraphene}, were we present several fitting curves obtained by sequentially disregarding the binding energies from smaller unit cells. Indeed, the asymptotic relation Eq.~(8) has a coefficient of determination, $r^2$, to the DFT binding energies bigger than 0.99 for the fitting that considers only unit-cell sizes bigger than 2 nm. We utilize this criteria in several systems and report the obtained fitting parameters in Table \ref{tab:Fitpar}. We also present in Fig.~\ref{fig:fig2Sup} the binding energies as function of box size in terms of the substrate rather than on the metal atom.

\vspace{-0.2cm}
  \begin{table}[h]
  \begin{center}
  \begin{tabular}{|c|c|c|c|c|} \hline
    color& $m$ & r$^2$ & B (eV nm$^2$) & $\Delta E (\infty)$ (eV) \\ \hline \hline
    \textcolor{red}{red}    &8&0.788&-0.0367 $\pm$ 0.009&-1.502 $\pm$ 0.005\\ \hline
    \textcolor{blue}{blue}  &7&0.878&-0.0721 $\pm$ 0.021&-1.495 $\pm$ 0.006\\ \hline
    \textcolor{green}{green}&6&0.971&-0.1168 $\pm$ 0.043&-1.488 $\pm$ 0.008\\ \hline
    \textcolor{brown}{brown}&5&0.998&-0.1492 $\pm$ 0.082&-1.485 $\pm$ 0.013\\ \hline    
  \end{tabular}
\end{center}
 \vspace{-0.3cm}
  \caption{{\bf Fittings for Al/(8,0) CNT}. Fitting parameters $B$ and $\Delta E (\infty)$, as well as the coefficient of determination $r^2$, for the curves in Fig. \ref{fig:fitAl}. The fitting parameters are reported within a standard error that we obtain by assuming uniform variance scale estimator and weights proportional to the DFT error of 0.01 eV. \label{tab:Alfit}}
%
  \begin{center}
  \begin{tabular}{|c|c|c|c|c|} \hline
    color& $m$ & r$^2$ & B (eV nm$^2$) & $\Delta E (\infty)$ (eV) \\ \hline \hline
    \textcolor{red}{red}    &8&0.903& 0.143 $\pm$ 0.007&-1.300 $\pm$ 0.005\\ \hline
    \textcolor{blue}{blue}  &7&0.935&-0.150 $\pm$ 0.013&-1.282 $\pm$ 0.007\\ \hline
    \textcolor{green}{green}&6&0.996&-0.172 $\pm$ 0.024&-1.289 $\pm$ 0.009\\ \hline
    \textcolor{brown}{brown}&5&0.999&-0.181 $\pm$ 0.043&-1.291 $\pm$ 0.013\\ \hline    
  \end{tabular}
\end{center}
 \vspace{-0.3cm}
  \caption{{\bf Fittings for Al/Graphene}. Fitting parameters $B$ and $\Delta E (\infty)$, as well as the coefficient of determination $r^2$, for the curves in Fig. \ref{fig:fitAlgraphene}. The fitting parameters are reported within a standard error that we obtain by assuming uniform variance scale estimator and weights proportional to the DFT error of 0.01 eV. \label{tab:Alfitgraphene}}
  \end{table}

\clearpage

\onecolumngrid
\section{Numerical Convergence}
In this section, we provide further details of the numerical accuracy of the binding energies reported here from DFT calculations.
\vspace{-0.5cm}
\onecolumngrid
\twocolumngrid


\vspace{0.2cm}
\begin{table}[H]
  \centering
  \begin{tabular}{|c|c|c|c|c|} \hline
    System& B &$\sigma_{\rm B}$ & $\Delta E (\infty)$ &$\sigma_{\Delta E (\infty)}$  \\
    &(eV nm$^2$)&(eV nm$^2$)& (eV)&(eV)\\\hline \hline
          Al Graphene&-0.1807&0.0431&-1.291&0.013 \\ \hline
          Al (5,5) CNT&-0.0564&0.0431&-1.427&0.013 \\ \hline
          Al (8,0) CNT&-0.1492&0.0433&-1.485&0.008 \\ \hline
          Ti (8,0) CNT&-0.0730&0.024&-2.375&0.008 \\ \hline
    V (8,0) CNT&0.0405&0.0227&-1.808&0.007\\ \hline
    Ni Graphene&0.0369&0.0068&-1.789&0.006  \\ \hline
  \end{tabular}
  \caption{{\bf Fitting parameters for Fig.\ 2}.  Fittings parameters $B$ and $\Delta E (\infty)$, and their corresponding standard errors, $\sigma_{\rm B}$ and $\sigma_{\Delta E (\infty)}$, for the fitting curves reported in Fig.~2 in the main text. \label{tab:Fitpar}}
\end{table}

\begin{center}
\begin{figure}[H]
  \centering
  \includegraphics[scale=0.3,trim=0 0 0 40,clip]{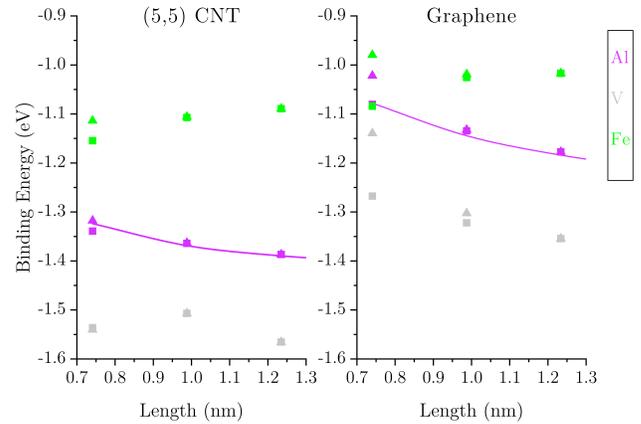}
  \vspace{-0.7cm}
  \caption{ {\bf Binding energies versus system size}. We show a close up of the variation in binding strength at shorter (5,5) CNTs (left) and smaller graphene sheets (right), for Al, V and Fe. The cell size varies from 0.7 nm to 1.3 nm. \label{fig:figZoomIn}}
\end{figure}
\end{center}

%
\vspace{-1.5cm}
\begin{center}
\begin{table}[H]
\begin{centering}
\begin{tabular}{|c|c|c|c|c|c|c|} \hline
\multicolumn{7}{|c|}{\bf 3$\times$3 supercell of Graphene } \\ \hline
$E_{\rm cut-off}$ (eV) &Al&Ti&Cu&Fe&Ni&V \\ \hline
400	&-1.062	&-2.141	&-0.484	&-1.037	&-1.700	&-1.265 \\ \hline
450	&-1.064	&-2.143	&-0.486	&-1.039	&-1.701	&-1.266 \\ \hline
500	&-1.064	&-2.143	&-0.486	&-1.040	&-1.702	&-1.267 \\ \hline
550	&-1.064	&-2.143	&-0.487	&-1.040	&-1.702	&-1.267 \\ \hline
600	&-1.065	&-2.143	&-0.487	&-1.040	&-1.702	&-1.267 \\ \hline \hline
\multicolumn{7}{|c|}{\bf 1$\times$1$\times$3 supercell of (5,5) BNNT } \\ \hline
$E_{\rm cut-off}$ (eV) &Al&Ti&Cu&Fe&Ni&V \\ \hline
400	&-0.623	&-1.274	&-0.523	&-0.559	&-1.492	&-0.908 \\ \hline
450	&-0.623	&-1.275	&-0.522	&-0.560	&-1.492	&-0.909 \\ \hline
500	&-0.624	&-1.275	&-0.523	&-0.561	&-1.493	&-0.910 \\ \hline
550	&-0.622 &-1.273	&-0.521	&-0.559	&-1.490	&-0.908 \\ \hline
600	&-0.622	&-1.273	&-0.521	&-0.559	&-1.491	&-0.908 \\ \hline 
\end{tabular}
\par\end{centering}
\caption{{\bf Convergence of the energy cut-off $E_{\rm cut-off}$.}  Convergence of $E_{\rm cut-off}$ of Al, Ti, Cu, Fe, Ni and V adsorbing to a 3$\times$3 supercell of graphene (top) and the 1$\times$1$\times$3 supercell of (5,5) BNNT (bottom). The convergence is tested through the binding strength of each metal atom to the substrate. Although an energy cut-off of 400 eV looks sufficient, due to the nature of our investigation we opted for an 450 eV cut-off. \label{tab:Encut}}
\end{table}
\end{center}
\vspace{1cm}
\begin{center}
\begin{table}[H]
\begin{centering}
\begin{tabular}{|c|c|c|c|c|c|c|} \hline
k-mesh&Al&Ti&Cu&Fe&Ni&V \\ \hline \hline
1$\times$1$\times$1	&-1.672	&-4.040	&-0.939	&-2.270	&-2.323	&-3.484 \\ \hline
2$\times$2$\times$1	&-0.671	&-1.716	&-0.483	&-1.028	&-1.698	&-1.114 \\ \hline
3$\times$3$\times$1	&-1.077	&-2.073	&-0.611	&-1.201	&-1.770	&-1.468 \\ \hline
4$\times$4$\times$1	&-1.071	&-1.921	&-0.493	&-1.069	&-1.731	&-1.258 \\ \hline
5$\times$5$\times$1	&-1.075	&-1.948	&-0.556	&-1.112	&-1.731	&-1.265 \\ \hline
6$\times$6$\times$1	&-1.045	&-1.916	&-0.500	&-1.075	&-1.726	&-1.262 \\ \hline
7$\times$7$\times$1	&-1.074	&-1.942	&-0.532	&-1.092	&-1.724	&-1.280 \\ \hline
8$\times$8$\times$1	&-1.084	&-1.929	&-0.507	&-1.076	&-1.721	&-1.256 \\ \hline
9$\times$9$\times$1	&-1.068	&-1.929	&-0.515	&-1.086	&-1.723	&-1.264 \\ \hline
10$\times$10$\times$1	&-1.069	&-1.924	&-0.509	&-1.077	&-1.720	&-1.261 \\ \hline
\textcolor{blue}{11$\times$11$\times$1}   &\textcolor{blue}{-1.080} &\textcolor{blue}{-1.932}	&\textcolor{blue}{-0.510}&\textcolor{blue}{-1.082}	&\textcolor{blue}{-1.723}	&\textcolor{blue}{-1.263} \\ \hline
12$\times$12$\times$1	&-1.080	&-1.926	&-0.509	&-1.077	&-1.719	&-1.258 \\ \hline
,13$\times$13$\times$1	&-1.075	&-1.928	&-0.510	&-1.080	&-1.722	&-1.263 \\ \hline
14$\times$14$\times$1	&-1.075	&-1.926	&-0.509	&-1.077	&-1.719	&-1.260 \\ \hline
15$\times$15$\times$1	&-1.078	&-1.928	&-0.510	&-1.079	&-1.721	&-1.260 \\ \hline
\end{tabular}
\par\end{centering}
\caption{{\bf Monkhorst Pack mesh convergence test in graphene.} The test is performed for
Al, Ti, Cu, Fe, Ni and V on the 3$\times$3 supercell of graphene. Convergence is tested through the binding strength of each metal atom to the substrate, reported in eV. As a result, we conclude that for this substrate, a $k$-point mesh of 11$\times$11$\times$1 is required to obtain converged results in the binding energies in all cases. \label{tab:kpointG}}
\end{table}
\end{center}

\vspace{2cm}
\twocolumngrid
\begin{center}
\begin{table}[H]
\begin{centering}
\begin{tabular}{|c|c|c|c|c|c|c|} \hline
k-mesh&Al&Ti&Cu&Fe&Ni&V \\ \hline \hline
1$\times$1$\times$1 &  -0.806	&-2.080	&-0.709	&-0.993	&-2.09	&-1.432 \\ \hline
1$\times$1$\times$2 &  -1.524	&-2.464	&-0.823	&-1.204	&-2.19	&-1.772 \\ \hline
1$\times$1$\times$3 &  -1.305	&-2.201	&-0.668	&-0.970	&-1.96	&-1.516 \\ \hline
1$\times$1$\times$4 &  -1.185	&-2.171	&-0.695	&-0.995	&-2.03	&-1.442 \\ \hline
1$\times$1$\times$5 &  -1.266	&-2.231	&-0.724	&-1.032	&-2.04	&-1.434 \\ \hline
1$\times$1$\times$6 &  -1.370	&-2.329	&-0.772	&-1.107	&-2.07	&-1.545 \\ \hline
1$\times$1$\times$7 &  -1.348	&-2.289	&-0.757	&-1.082	&-2.02	&-1.521 \\ \hline
1$\times$1$\times$8 &  -1.329	&-2.240	&-0.720	&-1.035	&-2.01	&-1.473 \\ \hline
1$\times$1$\times$9 &  -1.332	&-2.253	&-0.731	&-1.044	&-2.03	&-1.484 \\ \hline
\textcolor{blue}{1$\times$1$\times$10}&	\textcolor{blue}{-1.343}	&\textcolor{blue}{-2.271}	&\textcolor{blue}{-0.752}	&\textcolor{blue}{-1.068}	&\textcolor{blue}{-2.04}	&\textcolor{blue}{-1.502} \\ \hline
1$\times$1$\times$11&	-1.345	&-2.271	&-0.751	&-1.069	&-2.04	&-1.501 \\ \hline
1$\times$1$\times$12&	-1.339	&-2.266	&-0.745	&-1.062	&-2.03	&-1.496 \\ \hline
1$\times$1$\times$13&	-1.338	&-2.265	&-0.744	&-1.061	&-2.03	&-1.496 \\ \hline
1$\times$1$\times$14&	-1.339	&-2.266	&-0.745	&-1.062	&-2.04	&-1.497 \\ \hline
1$\times$1$\times$15&	-1.339	&-2.266	&-0.745	&-1.062	&-2.03	&-1.497 \\ \hline
\end{tabular}
\par\end{centering}
\caption{ {\bf Monkhorst Pack mesh convergence test in (5,5) CNT.} The test is performed for
Al, Ti, Cu, Fe, Ni and V on the 1$\times$1$\times$4 supercell of (5,5) CNT. Convergence is tested through the binding strength of each metal atom to the substrate, reported in eV. As a result, we conclude that for this substrate, a $k$-point mesh of 1$\times$1$\times$10 is required to obtain converged results in the binding energies in all cases. \label{tab:kpointCNT55}}
\end{table}
\end{center}
\vspace{-2cm}
%
\begin{center}
\begin{table}[H]
\begin{centering}
\begin{tabular}{|c|c|c|c|c|c|c|} \hline
k-mesh&Al&Ti&Cu&Fe&Ni&V \\ \hline \hline
1$\times$1$\times$1	&-1.858	&-3.299	&-1.161	&-1.612	&-2.247	&-2.428 \\ \hline
1$\times$1$\times$2	&-1.282	&-2.406	&-0.794	&-1.175	&-2.045	&-1.727 \\ \hline
1$\times$1$\times$3	&-1.541	&-2.457 &-0.913	&-1.221	&-2.051	&-1.769 \\ \hline
1$\times$1$\times$4	&-1.553	&-2.448	&-0.854	&-1.213	&-2.058	&-1.716 \\ \hline
1$\times$1$\times$5	&-1.505	&\textcolor{blue}{-2.453}&\textcolor{blue}{-0.889}	&\textcolor{blue}{-1.215} &\textcolor{blue}{-2.055} &-1.750 \\ \hline
1$\times$1$\times$6	&-1.506	&-2.447	&-0.878	&-1.214 &-2.056	&-1.740 \\ \hline
1$\times$1$\times$7	&\textcolor{blue}{-1.528}	&-2.453	&-0.881	&-1.214	&-2.055	&-1.754 \\ \hline
1$\times$1$\times$8	&-1.532	&-2.449	&-0.883	&-1.214	&-2.056	&\textcolor{blue}{-1.763} \\ \hline
1$\times$1$\times$9	&-1.518	&-2.453	&-0.878	&-1.215	&-2.056	&-1.764 \\ \hline
1$\times$1$\times$10	&-1.517	&-2.450	&-0.883 &-1.215	&-2.056	&-1.763 \\ \hline
1$\times$1$\times$11	&-1.528	&-2.453	&-0.881	&-1.215	&-2.056	&-1.763 \\ \hline
1$\times$1$\times$12	&-1.529	&-2.451	&-0.881	&-1.215	&-2.056	&-1.764 \\ \hline
1$\times$1$\times$13	&-1.522 &-2.453	&-0.882	&-1.215	&-2.056	&-1.764 \\ \hline
1$\times$1$\times$14	&-1.522 &-2.452	&-0.880	&-1.215	&-2.056	&-1.763 \\ \hline
1$\times$1$\times$15	&-1.528	&-2.453	&-0.883	&-1.215	&-2.056	&-1.764 \\ \hline
\end{tabular}
\par\end{centering}
\caption{{\bf Monkhorst Pack mesh convergence test in (8,0) CNT.} The test is performed for
Al, Ti, Cu, Fe, Ni and V on the 1$\times$1$\times$3 supercell of (8,0) CNT. Convergence is tested through the binding strength of each metal atom to the substrate, reported in eV. As a result, we conclude that for this substrate, a $k$-point mesh of 1$\times$1$\times$5 is required to obtain converged results in the binding energies in the cases of Ti, Cu, Fe, and Ni; a k-point mesh of 1$\times$1$\times$7 in the case of Al, and 1$\times$1$\times$8 for V.\label{tab:kpointCNT55}}
%
\end{table}
\end{center}

\begin{table}[H]
\begin{centering}
\begin{tabular}{|c|c|c|c|c|c|c|} \hline
k-mesh&Al&Ti&Cu&Fe&Ni&V \\ \hline \hline
1$\times$1$\times$1	&-0.635	&-1.273	&-0.540	&-0.564	&-1.509	&-0.924 \\ \hline
\textcolor{blue}{1$\times$1$\times$2}	&\textcolor{blue}{-0.623}	&\textcolor{blue}{-1.274}	&\textcolor{blue}{-0.522}	&\textcolor{blue}{-0.560}	&\textcolor{blue}{-1.491}	&\textcolor{blue}{-0.904} \\ \hline
1$\times$1$\times$3	&-0.623	&-1.273	&-0.522	&-0.559	&-1.492	&-0.903 \\ \hline
1$\times$1$\times$4	&-0.623	&-1.273	&-0.522	&-0.559	&-1.492	&-0.903 \\ \hline
1$\times$1$\times$5	&-0.623	&-1.273	&-0.522	&-0.559	&-1.492	&-0.903 \\ \hline
1$\times$1$\times$6	&-0.623	&-1.273	&-0.522	&-0.559	&-1.492	&-0.903 \\ \hline
\end{tabular}
\par\end{centering}
\caption{ {\bf Monkhorst Pack mesh convergence test in (5,5) BNNT.} The test is performed for
Al, Ti, Cu, Fe, Ni and V on the 1$\times$1$\times$3 supercell of (5,5) BNNT. Convergence is tested through the binding strength of each metal atom to the substrate, reported in eV. As a result, we conclude that for this substrate, a $k$-point mesh of 1$\times$1$\times$2 is required to obtain converged results in the binding energies in all cases. \label{tab:kpointCNT55}\label{tab:kpointBNNT55}}
\end{table}

\begin{center}
\begin{table}[H]
\begin{centering}
\begin{tabular}{|c|c|c|c|} \hline
System & 1 nm& 2 nm & Energy Difference \\ 
& (eV)&(eV)&(eV) \\ \hline \hline
clean surface & -297.446& -297.424&-0.022 \\ \hline
Fe  & -300.796&-301.168& -0.372\\ \hline
Fe (fixed) & -301.746& -301.709&-0.037\\ \hline
Ni  &-299.722&-299.691&-0.031 \\ \hline
Al  &-297.972&-297.940&-0.033 \\ \hline
Cu  &-298.168&-298.122&-0.046 \\ \hline
Ti  &-301.334&-301.291&-0.043 \\ \hline
\end{tabular}
\par\end{centering}
\caption{{\bf Layer spacing test using the 4$\times$4 Graphene supercell}. The total energies of Al, Ti, Fe, Ni, and Cu are presented for a 1 nm and 2 nm layer spacing. Note that in the case of Fe, the $T$ site (fixed) is tested next to the $H$ site since it presents a local minimum, with the $H$ site being the preferred adsorption site. The term ``fixed'' in the case of Fe indicates that Fe was constrained at this position. All other adsorption cases took place at the $H$ site.\label{tab:spacing}}
\end{table}
\end{center}

\vspace{-1.5cm}
\begin{figure}[H]
  \centering
  \includegraphics[scale=0.31]{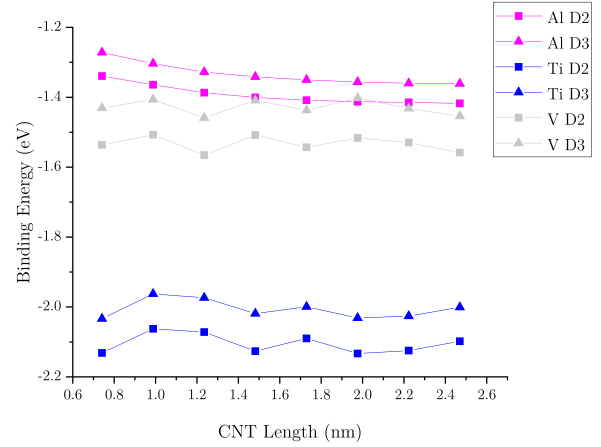}
  \includegraphics[scale=0.31]{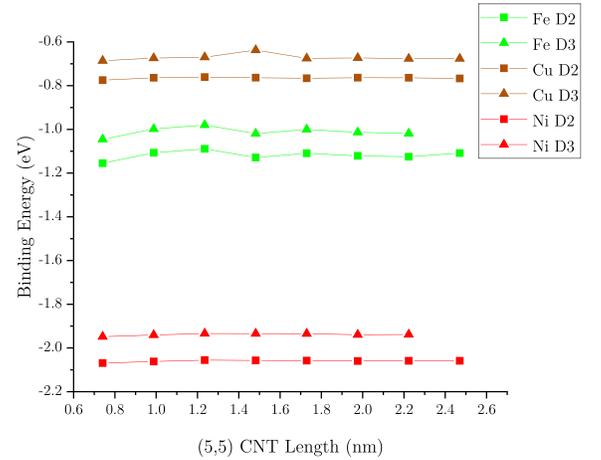} 
 \caption{{\bf Binding energies obtained from two different dispersion correction methods.} Binding energies D2 versus D3 (Grimme) of Al, Ti
and V (top) and Fe, Ni and Cu (bottom) adsorbing to the (5,5) CNT of different length.}\label{fig:D2vsD3}
\end{figure}